\documentclass[nofootinbib, showkeys, unsortedaddress, superscriptaddress, notitlepage,longbibliography, twocolumn]{revtex4-1}

\setcitestyle{authoryear,round}
\usepackage{etoolbox}
\makeatletter
\renewcommand\@biblabel[1]{\hspace*{\labelwidth}}
\apptocmd{\NAT@thebibliography}{\setlength\itemindent{-8pt}}{}{}
\makeatother

\makeatletter
\patchcmd{\NAT@citex}
  {\@citea\NAT@hyper@{%
     \NAT@nmfmt{\NAT@nm}%
     \hyper@natlinkbreak{\NAT@aysep\NAT@spacechar}{\@citeb\@extra@b@citeb}%
     \NAT@date}}
  {\@citea\NAT@nmfmt{\NAT@nm}%
   \NAT@aysep\NAT@spacechar\NAT@hyper@{\NAT@date}}{}{}

\patchcmd{\NAT@citex}
  {\@citea\NAT@hyper@{%
     \NAT@nmfmt{\NAT@nm}%
     \hyper@natlinkbreak{\NAT@spacechar\NAT@@open\if*#1*\else#1\NAT@spacechar\fi}%
       {\@citeb\@extra@b@citeb}%
     \NAT@date}}
  {\@citea\NAT@nmfmt{\NAT@nm}%
   \NAT@spacechar\NAT@@open\if*#1*\else#1\NAT@spacechar\fi\NAT@hyper@{\NAT@date}}
  {}{}
  
\makeatother

\usepackage{graphicx}
\usepackage{subcaption}
\usepackage{amsmath}
\usepackage{amssymb}
\usepackage{dsfont}
\usepackage{mathptmx}
\usepackage{geometry}
\usepackage{xcolor}
\usepackage{MnSymbol}
\usepackage{enumitem}
\usepackage{mathtools}

\newcommand{\EE}{\mathbb{E}}
\newcommand{\RR}{\mathbb{R}}

\newcommand{\NN}{\mathbb{N}}
\newcommand{\PP}{\mathbb{P}}
\newcommand{\eps}{\varepsilon}
\DeclareMathOperator*{\argmin}{arg\,min}
\DeclareMathOperator{\Det}{Det}
\DeclareMathOperator{\ppr}{pr}
\DeclareMathOperator{\Id}{Id}
\newcommand{\dd}{\mathrm{d}}
\DeclareMathOperator{\rank}{rank}
\DeclareMathOperator{\trace}{tr}
\newcommand{\dv}[2]{\frac{\dd#1}{\dd#2}}

\newcommand{\fdv}[2]{\frac{\delta #1}{\delta #2}}
\newcommand{\nfdv}[3]{\frac{\delta^{#1}#2}{\delta#3^{#1}}}
\newcommand{\abs}[1]{\left\lvert#1\right\rvert}
\newcommand{\norm}[1]{\left\lVert#1\right\rVert}

\newcommand{\figEnd}{eps}

\makeatletter
\renewcommand\@makecaption[2]{%
  \par
  \vskip\abovecaptionskip
  \begingroup
   \small\rmfamily
    \begingroup
     \samepage
     \flushing
     \let\footnote\@footnotemark@gobble
     \@make@capt@title{#1}{#2}\par
    \endgroup
  \endgroup
  \vskip\belowcaptionskip
}
\makeatother

\makeatletter
\renewcommand{\p@subsection}{}
\renewcommand{\p@subsubsection}{}
\makeatother

\usepackage{chngcntr}
\usepackage[linktocpage,colorlinks=true,allcolors=blue]{hyperref}

\begin{document}

\title{Scalable Methods for Computing Sharp Extreme Event Probabilities in\\
 Infinite-Dimensional Stochastic Systems}

\author{Timo Schorlepp}
\email{Timo.Schorlepp@rub.de}
\affiliation{Institute for Theoretical Physics I, Ruhr University Bochum, Bochum, Germany}
\author{Shanyin Tong}
\email{st3503@columbia.edu}
\affiliation{Department of Applied Physics and Applied Mathematics, Columbia University, New York, NY, USA}
\author{Tobias Grafke}
\email{T.Grafke@warwick.ac.uk}
\affiliation{Mathematics Institute, University of Warwick, Coventry, United Kingdom}
\author{Georg Stadler}
\email{stadler@cims.nyu.edu}
\affiliation{Courant Institute of Mathematical Sciences, New York University, New York, NY, USA}

\date{\today}

\begin{abstract}
  We introduce and compare computational techniques for sharp extreme
  event probability estimates in stochastic differential equations
  with small additive Gaussian noise. In particular, we focus on
  strategies that are scalable, i.e.\ their efficiency does not
  degrade upon temporal and possibly spatial refinement.
  For that purpose, we extend
  algorithms based on the Laplace method for estimating the
  probability of an extreme event to infinite dimensional path space.
  The method
  estimates the limiting exponential scaling using a single
  realization of the random variable, the large deviation
  minimizer. Finding this minimizer amounts to solving an optimization
  problem governed by a differential equation.  The probability
  estimate becomes sharp when it additionally includes prefactor
  information, which necessitates computing the determinant of a
  second derivative operator to evaluate a Gaussian integral around the minimizer. We
  present an approach in infinite dimensions based on
 Fredholm determinants, and develop numerical algorithms to compute
 these determinants efficiently for the high-dimensional systems that
 arise upon discretization. We also give an interpretation of this
 approach using Gaussian process covariances and transition tubes.
 An example model problem, for which we provide an open-source
 python implementation, is used throughout the paper to illustrate all
 methods discussed.
 To study the performance of the methods, we consider examples of
 stochastic differential and stochastic partial differential
 equations, including the randomly forced incompressible
 three-dimensional Navier--Stokes equations.

  \keywords{stochastic differential equations, extreme events, large
    deviation theory, precise Laplace asymptotics, Fredholm
    determinant, Navier--Stokes equations}
\end{abstract}

\maketitle

\tableofcontents

\section{Introduction}

The estimation of extreme event probabilities in complex stochastic
systems is an important problem in applied sciences and engineering,
and is difficult as soon as these events are too rare to be easily
observable, but at the same time too impactful to be ignored.
Examples of such events studied in the recent literature include rogue
waves~\citep{dematteis-grafke-vanden-eijnden:2018} and
wave impacts on an offshore platform~\citep{mohamad-sapsis:2018},
heat waves and cold spells~\citep{ragone-wouters-bouchet:2018,galfi-lucarini-wouters:2019},
intermittent fluctuations in turbulent flows~\citep{fuchs-herbert-rolland:2022}
and derivative pricing fluctuations in mathematical finance~\citep{friz-gatheral-gulisashvili:2015}. A broad perspective on extreme event
prediction can be found in~\citet{farazmand-sapsis:2019}. Methods to estimate extreme events typically rely on Monte Carlo simulations, including importance sampling \citep{bucklew:2013}, subset simulation \citep{au-beck:2001} or multilevel splitting methods \citep{budhiraja-dupius:2019}.

A possible theoretical framework to assess extreme event
probabilities, which we will follow in this work, is given by large
deviation theory (LDT) \citep{varadhan:1984,dembo-zeitouni:1998}. This
approach allows to estimate the dominant, exponential scaling of the
probabilities in question through the solution of a deterministic
optimization problem, namely finding the most relevant realization of
the stochastic process for a given outcome. This realization is
sometimes called instanton, inspired by theoretical physics.  For
stochastic processes described by stochastic differential equations
(SDEs), the relevant theory has been formulated
by~\citet{freidlin-wentzell:2012}, and can be extended to many
stochastic partial differential equations (SPDEs).  The computational
potential of this formulation has been reviewed
by~\citet{grafke-vanden-eijnden:2019}.

In addition to the exponential scaling provided by LDT,
it is often desirable to obtain asymptotically sharp, i.e.\ asymptotically
exact probability estimates. This requires
the evaluation of a pre-exponential factor in addition to the
usual leading-order large deviation result, when interpreting LDT as a
Laplace approximation. On the theoretical side, there exist multiple
results for such precise Laplace asymptotics for general SDEs~\citep{ellis-rosen:1982,azencott:1982,ben-arous:1988,piterbarg-fatalov:1995,deuschel-etal:2014}
and certain SPDEs requiring
renormalization~\citep{berglund-di_gesu-weber:2017,friz-klose:2022},
which, however, typically do not include an actual evaluation
of the abstract objects in terms of which they are formulated.
We concentrate on the case of SDEs or well-posed SPDEs with additive noise
here, where computing the leading-order prefactor amounts
to evaluating a Fredholm determinant of an integral operator.

\begin{figure}
  \centering
  \includegraphics[width = .425 \textwidth]{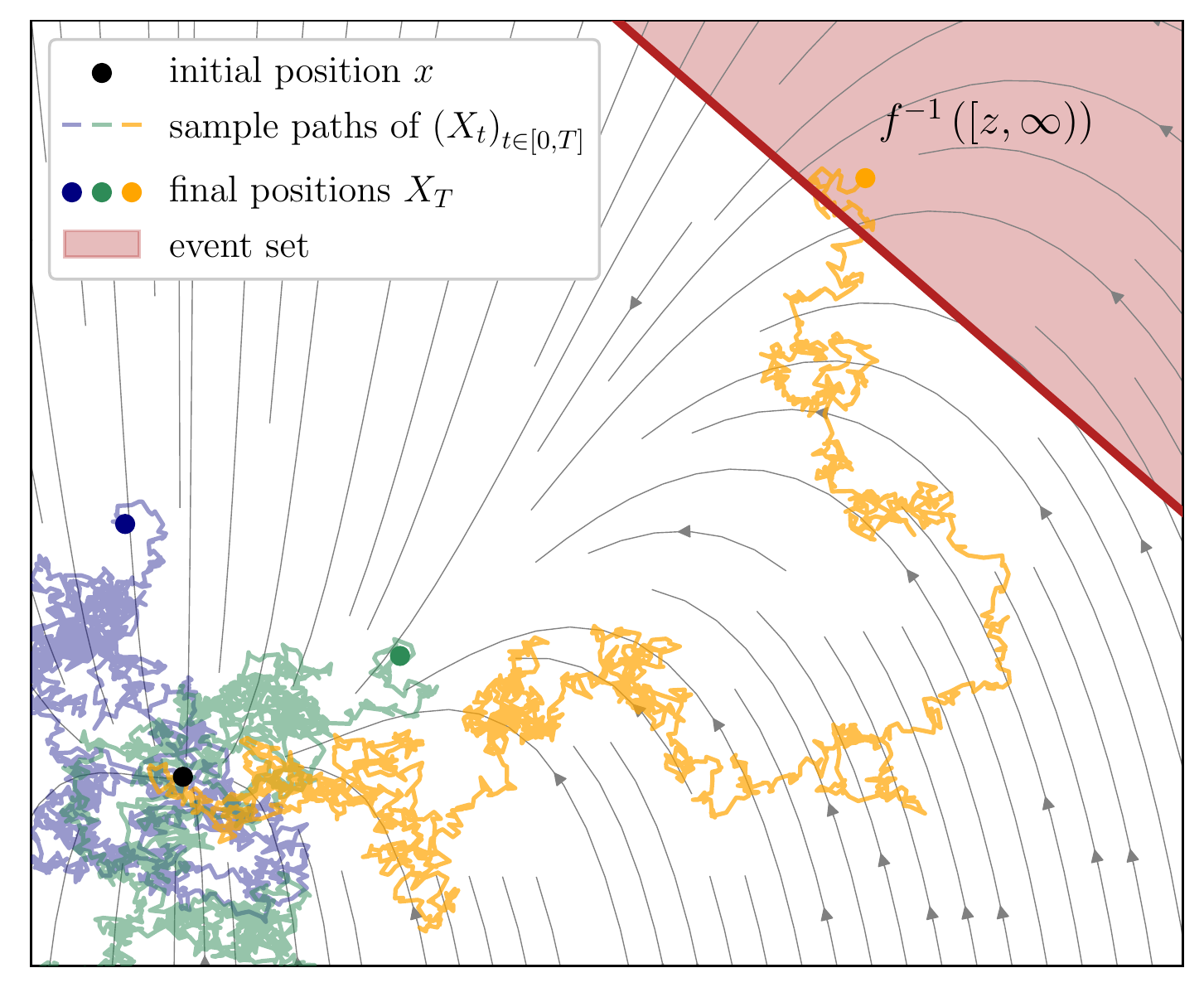}
  \caption{Visualization of extreme event set (red), a sample path
    that, from a given initial condition, ends in the extreme event set
    at final time $T$ (orange) and two typical sample paths that
    do not end
    in the event set (blue and green). The gray lines are field lines of
    the drift vector field~$b$. This is the reason why paths that
    end in the event set are rare,
    since the noise must act against the flow of the deterministic
    vector field $b$ to
    push the system into the extreme event set.
    Details of this example problem, which
    is used throughout the paper as illustration, will be given in Section
    \ref{subsec:laplace-infinite} and more rare event paths and the
    instanton are shown in Figure~\ref{fig:2mb-paths}.
    The implementation of this
    example is available from a public GitHub repository
    \citep{Schorlepp-github}.}
\label{fig:2mb-sketch}
\end{figure}

\paragraph*{Approach.}
In this paper, we present a sharp and computable probability estimate
for tail
probabilities $\PP \left[ f\left(X_T\right) \geq z \right]$,
i.e.\ a real-valued function $f$ of a diffusion
process $(X_t)_{t \in [0,T]}$ with state space $\RR^n$ and
\begin{align}
\begin{cases}
\dd X_t = b(X_t) \dd t + \sigma \dd B_t\,,\\
X_0 = x \in \RR^n,
\end{cases}
\label{eq:sde-intro}
\end{align}
exceeding a given threshold $z$ at final time $T$ (see
Figure~\ref{fig:2mb-sketch} for an example of this setup).
We demonstrate that
\begin{align}
\PP \left[ f\left(X_T\right) \geq z \right] \approx (2 \pi)^{-1/2}
C(z) \exp \left\{-I(z) \right\}\,,
\end{align}
in a way to be made precise later on, with real-valued
functions $C$, called the (leading-order) prefactor, and $I$,
called the rate function. The latter is determined through the
solution of a constrained optimization problem:
\begin{align}
  I(z) = \min_{\begin{subarray}{c}\eta\in L^2([0,T],\RR^n)\\
  \text{s.t. }f\left(X_T[\eta] \right) = z
    \end{subarray}}  \;
  \frac{1}{2}
\norm{\eta}_{L^2}^2,
\label{eq:if-intro}
\end{align}
where formally $\eta = \dd B_t / \dd t$ is the time derivative of
the Brownian motion $(B_t)_{t \in [0,T]}$, and $X_T$ depends on $\eta$
through~\eqref{eq:sde-intro}. The prefactor $C$ is then expressed
as a Fredholm determinant of a linear operator which contains the
solution of the minimization problem~\eqref{eq:if-intro},
the instanton $\eta_z$, as a
background field and acts on paths $\delta \eta \colon [0,T] \to \RR^n$.
We show how to evaluate this operator determinant numerically
for general SDEs and SPDEs, and demonstrate through multiple examples
that it is possible to do so even for very high-dimensional systems with $n \gg 1$ arising, for instance, after spatial discretization of an SPDE.
Our approach is based on computing
the dominant eigenvalues of the trace-class integral operator entering
the Fredholm determinant.

\paragraph*{Related literature.}
In the physics literature, the leading-order prefactor computation
corresponds to the evaluation of Gaussian path integrals,
which is a classical topic in
quantum and statistical field theory~\citep{zinn-justin:2021}.
There are multiple references dealing with the evaluation of such integrals
for the class of differential operators that is necessary for SDEs, such
as~\citet{papadopoulos:1975,nickelsen-engel:2011,corazza-fadel:2020}. In accordance with these approaches,
in the last years, numerical
leading order prefactor computation methods for general SDEs and SPDEs via the
solution of Riccati matrix differential equations have been established~\citep{schorlepp-grafke-grauer:2021,ferre-grafke:2021,grafke-schaefer-vanden-eijnden:2021,bouchet-reygner:2022,schorlepp-grafke-grauer:2023}.
An early example using a similar method is given
by~\citet{maier-stein:1996}. All of these papers have in common that
the leading order prefactor can be evaluated in a closed form by
solving a single matrix valued initial or final value problem,
thereby bypassing the need
to compute large operator determinants directly. We briefly introduce
this method in this paper, relate it to the -- in some sense
complementary -- Fredholm determinant prefactor
evaluation based on dominant eigenvalues, and discuss possible
advantages and disadvantages.
We note that for SDEs with low-dimensional state space, it can also be
feasible to compute
the differential operator determinants, that are otherwise evaluated
through the Riccati matrices, directly by discretizing
the operator into a large matrix and numerically calculating its
determinant, which has been carried out e.g.\
by~\citet{psaros-kougioumtzoglou:2020,zhao-psaros-petromichelakis-etal:2022}.

Another perspective on the precise Laplace approximation used
in this paper is provided
by the so-called second-order reliability method (SORM), which is used in
the engineering literature to estimate failure probabilities,
as reviewed e.g.\ by~\citet{rackwitz:2001,breitung:2006}.
For example, the asymptotic form of the extreme event
probabilities in this paper corresponds to the standard form stated
by~\citet{breitung:1984}. In this sense, the method proposed in this paper
can be regarded as a path space SORM, carried over to infinite dimensions
for the case of additive noise SDEs. The connection of precise LDT estimates
to SORM for finite-dimensional parameter spaces has also been pointed out
by~\citet{tong-vanden-eijnden-stadler:2021}.

In studies of rare and extreme event estimation, Monte Carlo
simulations are commonly used, and various sampling schemes have been
designed, some of which have been modified and adapted to systems
involving SDEs. These include various importance sampling estimators
which can be associated e.g.\ with the solution to deterministic
optimal control problems along random trajectories
\citep{vanden-weare:2012}, with the instanton in LDT
\citep{ebener-margazoglou-friedrich-etal:2019}, or build on stochastic
Koopman operator
eigenfunctions \citep{zhang-sahai-marzouk:2022}. The method we propose
takes a different perspective from these sampling methods---it does
not involve sampling, and is only asymptotically exact.

\paragraph*{Contributions and limitations.}
The main contributions of this paper are as follows:
(i) Generalizing SORM to infinite dimensions, we introduce a
  sampling-free method to approximate extreme event probabilities for
  SDEs (and SPDEs) with additive noise. The method is based on the
  Laplace approximation in path space and uses second-order
  information to compute the probability prefactor.
(ii) While such precise Laplace asymptotics for SDEs are known on a
  theoretical level, we show how to evaluate them numerically in a
  manner that is straightforward to implement and is scalable,
  i.e.\ it does not degrade with increasing discretization
  dimension. We illustrate the method on a high-dimensional nonlinear
  example, namely estimating the probability of high strain rate
  events in a three-dimensional stochastic Navier-Stokes flow.
(iii) On the theoretical level, we explore the relationship between
  the proposed eigenvalue-based approach for calculating the prefactor
  and Riccati methods from stochastic analysis and stochastic field
  theory.  We examine the advantages of each method
  and provide an interpretation of the
  involved Gaussian process using transition tubes
  towards the extreme event, i.e.\ the expected magnitude and
  direction of fluctuations on the way to an extreme outcome.

The approach taken in this paper also has some limitations:
(i)  While we find the probability estimates including the
  leading-order prefactor to be quite accurate when compared to direct
  Monte Carlo simulations when these are feasible, these estimates are
  approximations and only asymptotically exact in the limit as
  $z\to\infty$. To obtain unbiased estimates, one can e.g.\ use
  importance sampling. The instanton and the second variation
  eigenvalues and eigenvectors can be used as input for such extreme
  event importance sampling
  algorithms~\citep{ebener-margazoglou-friedrich-etal:2019,tong-vanden-eijnden-stadler:2021,tong-stadler:2022}.
(ii) We limit ourselves to SDEs with additive Gaussian noise.  For
  SDEs with multiplicative noise (or singular SPDEs), the
  leading-order prefactor is  more complicated, as the
  direct analogy to the finite-dimensional case gets
  lost~\citep{ben-arous:1988}. Nevertheless, extensions of the
  eigenvalue-based prefactor computation proposed here can likely be
  made, but are beyond the scope of this paper.
(iii) The proposed approach assumes that the differential
  equation-optimization \eqref{eq:if-intro} has a unique solution that
  can be computed. For non-convex constraints, uniqueness may be
  difficult to prove or may not hold. However, in the examples we
  consider, we seem to be able to identify the global minimizer
  reliably by using several different initializations in the
  minimization algorithm and, if we find different minimizers, by
  choosing the one corresponding to the smallest objective value.  The
  proposed approach can also be generalized to multiple isolated and
  continuous families of
  minimizers~\citep{ellis-rosen:1981,schorlepp-grafke-grauer:2023}.

\paragraph*{Notation.}
We use the following notations throughout the paper: The state space
dimension is always written as~$n$, a possible time discretization
dimension of the interval $[0,T]$ as~$n_t$, and~$N$ is exclusively used in
section~\ref{subsec:laplace-finite} for the motivation of our results
via random variables in $\RR^N$. We denote the
Euclidean norm and inner product in $\RR^N$ by $\norm{\cdot}_N$ and
$\langle \cdot, \cdot \rangle_N$, respectively, and the $L^2$ norm and
scalar product for $\RR^n$-valued functions defined on $[0,T]$
by $\norm{\cdot}_{L^2([0,T], \RR^n)}$ and $\langle \cdot, \cdot
\rangle_{L^2([0,T], \RR^n)}$, respectively. The outer product is denoted
by $\otimes$, with $v \otimes w = v w^\top$ and $v^{\otimes 2} \coloneq v \otimes v$ for $v,w \in \RR^N$ and $(f
\otimes g)(t,t') = f(t) g(t')^\top$ for $f,g \in L^2([0,T], \RR^n)$
and $t,t' \in [0,T]$. Convolutions are written as $*$.
The subscript or argument $z \in \RR$ always represents the dependency on
the observable value e.g.\ of the minimizer $\eta_z$, Lagrange
multiplier~$\lambda_z$
and projected second variation operator $A_z$, as well as the observable
rate function $I_F(z)$ and prefactor $C_F(z)$. The identity map is in
general denoted by $\Id$, and the identity matrix and zero matrix
in $\RR^N$ are written as $1_{N \times N}$ and $0_{N \times N}$.
The superscript $\perp$ always denotes the orthogonal complement,
with $v^\perp \coloneq (\text{span}(\{v\}))^\perp$.
Functional
derivatives with respect to $\eta \in L^2([0,T], \RR^n)$ are denoted
by $\delta / \delta \eta$. Determinants in $\RR^N$, as well as
Fredholm determinants, are written as $\det$, whereas regularized
differential operator determinants are written as $\Det$ with the
boundary conditions of the operator as a subscript. For two
real functions $g$ and $h$, we write
\begin{align}
g(\eps) \overset{\eps \downarrow 0}{
\sim} h(\eps) \quad \iff \quad \lim_{\eps \downarrow 0} \; \frac{g(\eps)}{h(\eps)} = 1\,, \label{eq:sharp-def}
\end{align}
if the functions $g$ and $h$ are asymptotically equivalent
as $\eps \downarrow 0$.
By an abuse of terminology, we use
the term ``instanton'' in this paper to refer to the large deviation
minimizer~$\eta_z$ for finite-dimensional parameter spaces,
and also to both the instanton noise trajectory
$\left(\eta_z(t)\right)_{t \in [0,T]}$ and the instanton
state variable trajectory $\left(\phi_z(t)\right)_{t \in [0,T]}$
in the infinite-dimensional setup.

We start with a more precise
explanation of the concepts described in this introduction in sections~\ref{subsec:laplace-finite}
and~\ref{subsec:laplace-infinite}, before summarizing the structure
of the rest of the paper at the end of section~\ref{subsec:laplace-infinite}.

\subsection{Laplace method for normal random variables in~$\RR^N$}
\label{subsec:laplace-finite}

We start with the finite dimensional setting,
following~\cite{dematteis-grafke-vanden-eijnden:2019,tong-vanden-eijnden-stadler:2021}:
We consider a collection of $N$ random parameters $\eta\in\RR^N$ that
are standard normally distributed, and are interested in a physical
\emph{observable}, described by a function $F\colon\RR^N\to\RR$, that
describes the outcome of an experiment under these random parameters.
Note that restricting ourselves to independent standard normal
variables is not a major limitation as $F$ may include a map that
transforms a standard normal to another distribution.
To give an example that fits into this setting, $\eta$ could
be all parameters entering a weather prediction model, and $F$ then
constitutes the mapping of the parameters to some final prediction,
such as the temperature at a given location in the future. Note
that
the map $F$ may be complicated and expensive to
evaluate, e.g.\ requiring the solution of a PDE.

We are interested in the probability that the outcome of the
experiment exceeds some threshold~$z$, i.e.~$P(z) = \PP[F(\eta) \geq
  z]$.  Since here $z$ is assumed large compared to typically expected
values of $F(\eta)$, we call~$P(z)$ the \emph{extreme event
  probability}.  To be able to control the rareness of the event, we
introduce a formal scaling parameter $\eps>0$ and consider $\eps\ll1$
to make the event extreme by defining $P_F^\eps(z) =
\PP[F(\sqrt{\eps}\eta) \geq z]$.  This allows us to treat terms of
different orders in $\eps$ perturbatively in the rareness of the event
and is more amenable to analysis than rareness due to $z\to\infty$. In
the following, we will thus consider $z$ as a fixed constant, while
discussing the limit $\eps\to0$. Since $\eta$ is normally distributed, the
extreme event probability is available as an integral,
\begin{equation}
  \label{eq:extreme-event-finite-dimensional}
  P_F^\eps(z) = (2\pi\eps)^{-N/2}\!\!\int_{\RR^N} \mathds{1}_{
  \{F(\eta)\geq z\}}(\eta) \exp\left\{-\frac1{2\eps}
  \norm{\eta}_N^2\right\}  \dd^N\eta\,,
\end{equation}
by integrating all possible $\eta$ that lead to an exceedance of the
observable threshold (as identified by the indicator function $\mathds{1}$),
weighed by their respective probabilities given by the Gaussian
densities. Directly evaluating the integral
in~\eqref{eq:extreme-event-finite-dimensional} is typically infeasible
for complicated sets $\{\eta\in \RR^N \mid F(\eta)\geq z\}$ and large
$N$.

The central notion of this paper is the fact that in the limit
$\eps\downarrow 0$, the integral
in~(\ref{eq:extreme-event-finite-dimensional}) can be approximated via the
\emph{Laplace method}, which replaces the integrand with its
extremal value, times higher order multiplicative corrections. The corrections at leading order in~$\eps$
amount to a Gaussian integral that can be
solved exactly. In effect, the
integral~(\ref{eq:extreme-event-finite-dimensional}) is approximated by
the probability of the \emph{most likely} event that exceeds the
threshold, multiplied by a factor that takes into account the event's neighborhood.

To make things concrete, we make the following assumptions on $F \in
C^2(\RR^N, \RR)$ for given $z > F(0)$:
\begin{enumerate}[leftmargin=*]
\item There is a unique $\eta_z \in \RR^N \backslash \{0\}$, called
the \textit{instanton}, that minimizes the function $\tfrac{1}{2} \norm{\cdot}^2_N$
in $F^{-1}([z,\infty))$. Necessarily,
$\eta_z \in F^{-1}(\{z\})$ lies on the boundary,
$F(\eta_z) = z$, and there
exists a Lagrange multiplier $\lambda_z\geq 0$ with $\eta_z =
\lambda_z \nabla F(\eta_z)$ as a first-order necessary condition.
We define the \textit{large deviation rate function} of the
family of real-valued random
variables~$\left( F(\sqrt{\eps} \eta) \right)_{\eps > 0}$ at $z$ via
\begin{align}
I_F \colon
\RR \to \RR\,, \quad I_F(z) := \tfrac{1}{2} \norm{\eta_z}^2_N\,.
\end{align}
\item $1_{N \times N} - \lambda_z \nabla^2 F(\eta_z)$ is positive
definite on the $(N-1)$-dimensional subspace $\eta_z^\perp \subset
\RR^N$ orthogonal to the instanton, i.e.\ we assume a second-order sufficient
condition for $\eta_z$ holds.
\end{enumerate}

Then, there is a sharp estimate, in the sense of~\eqref{eq:sharp-def},
for the extreme event
probability~\eqref{eq:extreme-event-finite-dimensional} via
\begin{equation}
  \label{eq:extreme-event-finite-dimensional-result}
  P_F^\eps(z) \overset{\eps\downarrow0}{\sim}\eps^{1/2} (2 \pi)^{-1/2}
  \, C_F(z) \,\exp\left\{-\frac1\eps I_F(z)\right\}\,,
\end{equation}
where the rate function $I_F$ determines the exponential scaling,
and $C_F(z)$ is the $z$-dependent leading order \emph{prefactor}
contribution that accounts for the local properties around the
instanton.
Note that
the pre\-factor is essential to get a sharp estimate, which cannot be
obtained from mere $\log$-asymptotics using only the rate function.
The prefactor $C_F(z)$ can explicitly be computed via
\begin{align}
C_F(z) = \left[2 I_F(z) \det \left(1_{N \times N} - \lambda_z
\ppr_{\eta_z^\perp} \nabla^2 F(\eta_z) \ppr_{
\eta_z^\perp} \right) \right]^{-1/2},
\label{eq:extreme-event-finite-dimensional-prefac}
\end{align}
where $\ppr_{\eta_z^\perp} = 1_{N \times N} - \eta_z \otimes
\eta_z / \norm{\eta_z}^2_N$ is the orthogonal projection onto
$\eta_z^\perp$. A brief derivation of this result, analogous
to the computations of~\cite{tong-vanden-eijnden-stadler:2021},
is included in appendix~\ref{sec:deriv}\ref{subsec:deriv-laplace-finite-dim} for
completeness. It is also directly equivalent to the standard form
of the second order reliability method, as derived e.g.\
by~\cite{breitung:1984}. Geometrically, it corresponds to replacing
the extreme event set $\{\eta \in \RR^N \mid F(\eta) \geq z \}$ by
a set bounded by
the paraboloid with vertex at the instanton~$\eta_z$, the axis of symmetry in the direction of $\nabla F(\eta_z)$ and curvatures
adjusted to be the eigenvalues of the $-\|\nabla F\|^{-1}$-weighted Hessian of $F$ at~$\eta_z$.

For the weather prediction example,
equations~(\ref{eq:extreme-event-finite-dimensional-result})
and~(\ref{eq:extreme-event-finite-dimensional-prefac}) mean the
following: We could
estimate~(\ref{eq:extreme-event-finite-dimensional}) by performing a
large number of simulations of the weather model with a random choice
of parameters to obtain statistics on an extremely high temperature
event. Instead, we solve an optimization problem over parameters to compute only the
single most likely route to that large temperature. When the desired
event is very extreme, such a situation can only be
realized when all simulated physical processes conspire in exactly the
right way to make the extreme temperature event
possible. Consequently, only a narrow choice of model parameters and
corresponding sequence of events remains that can contribute to the
extreme event probability: precisely the instanton singled out by the
optimization procedure. The probability of the extreme event
is then well approximated by perturbations around that single
most likely extreme outcome.

Next, we generalize the
statement~\eqref{eq:extreme-event-finite-dimensional-result} to the
infinite-dimensional setting encountered in continuous time stochastic
systems. Intuitively, for temporally evolving systems with stochastic
noise, there is randomness at every single instance in time, which
implies an infinite number of random parameters to optimize
over. We generalize the above strategy to the
important case of SDEs in $\RR^n$ driven by small additive Gaussian
noise, and assemble and compare computational methods to compute $I_F$
and $C_F$ numerically, even for very large spatial dimensions $n$
stemming from semi-discretizations of multi-dimensional SPDEs.

\subsection{Generalization to infinite dimensions for SDEs with additive noise}
\label{subsec:laplace-infinite}
As a stochastic model problem, we consider the SDE
\begin{align}
\begin{cases}
\dd X^\eps_t = b(X^\eps_t) \dd t + \sqrt{\eps} \sigma \dd B_t\,,\\
X^\eps_0 = x \in \RR^n,
\end{cases}
\label{eq:sde}
\end{align}
on the time interval $[0,T]$ with a deterministic initial condition and
$n \in \NN$, $\eps > 0$.  The drift vector field $b \colon \RR^n \to
\RR^n$, assumed to be smooth, may be nonlinear and non-gradient. The
constant matrix $\sigma \in \RR^{n \times n}$ is not required to be
diagonal or invertible. The SDE is driven by a standard
$n$-dimensional Brownian motion $B = (B_t)_{t \in [0,T]}$.  We limit
ourselves to the estimation of extreme event probabilities (due to
small noise $\epsilon$) of the
random variable $f(X^\eps_T)$, where $f\colon\RR^n \to \RR$ is a
smooth, possibly nonlinear observable of the process
$X^\eps$ at final time $t = T$.

\begin{figure}
  \centering
  \includegraphics[width = .48 \textwidth]{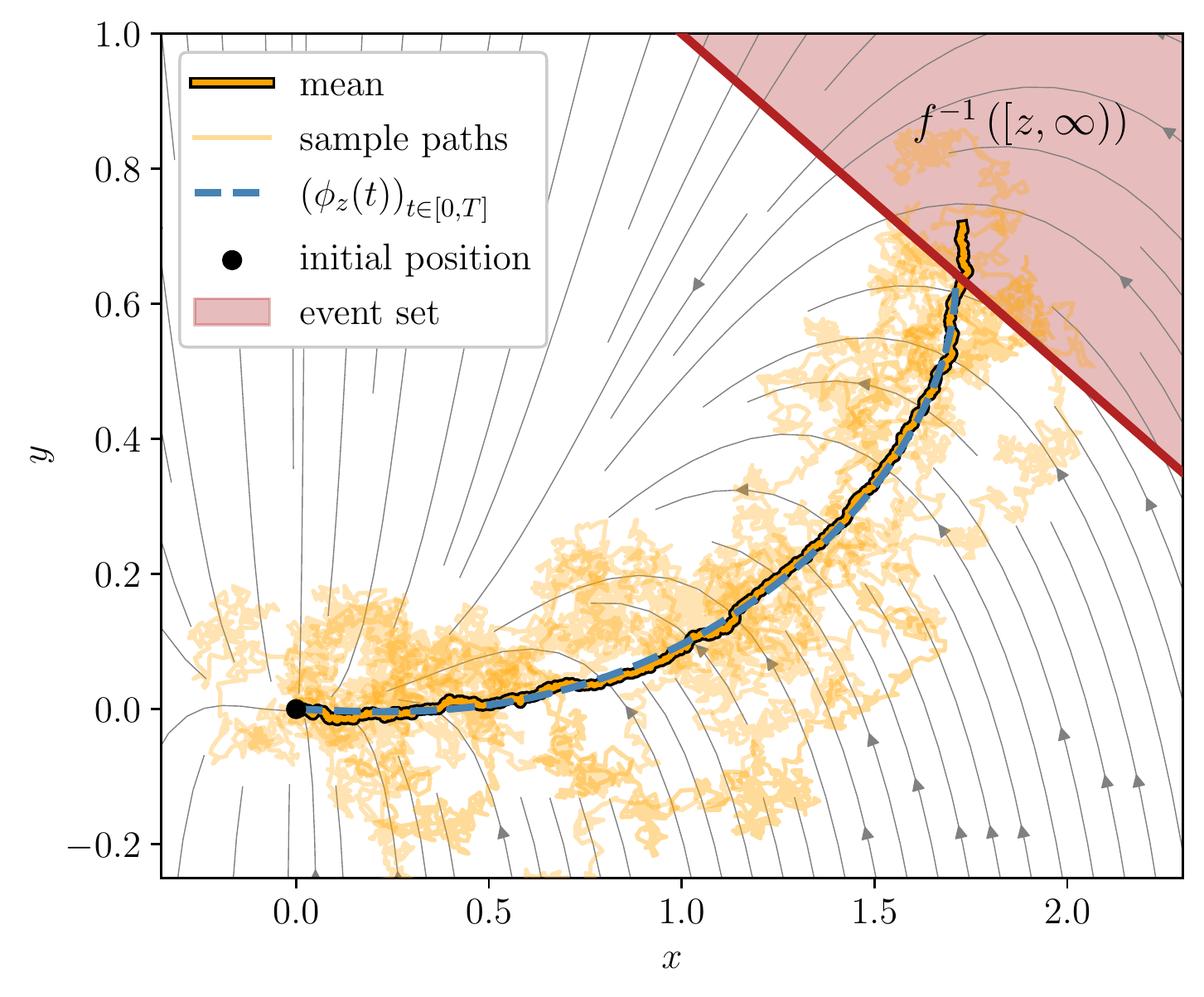}
  \caption{Visualization of five different sample paths (light orange)
    and the mean of 100 such paths (orange with black outline) of the
    model SDE~\eqref{eq:2d-example} that satisfy $f(X(T),Y(T)) \geq
    z$ with $z = 3$ (red set) and $\eps = 0.5$. Using Euler-Maruyama
    steps with an integrating factor with step size $\Delta t = 5
    \cdot 10^{-4}$, we repeatedly simulated~\eqref{eq:2d-example}
    until 100 such rare trajectories were found. The dashed blue line
    is the state variable instanton trajectory $\phi_z$, solution
    of~\eqref{eq:first-order-adj-eq} with the optimal $\eta_z$ as
    forcing. As in Figure~\ref{fig:2mb-sketch}, the gray lines are
    field lines of the drift vector field~$b$.}
\label{fig:2mb-paths}
\end{figure}

A concrete example of this type of system,
already alluded to in the first section, is shown in
figure~\ref{fig:2mb-paths}. It is given by the SDE
\begin{align}
  \label{eq:2d-example}
  &\begin{cases}
    \dd X = (-X-XY)\,\dd t + \sqrt{\eps}\,\dd B_X,\\
    \dd Y = (-4Y+X^2)\,\dd t + \tfrac12\sqrt{\eps}\,\dd B_Y,
  \end{cases} \nonumber\\
  &\text{with }(X(0),Y(0)) = (0,0)\,.
\end{align}
The streamlines in the figure show the motion taken by deterministic
trajectories of the model at $\eps = 0$.
Small magnitude stochasticity in the
form of Brownian noise is added, and we ask the question: What is the
probability $P_F^\eps(z)$, as defined below in~\eqref{eq:tail-def}, that the system ends up, at time $T=1$, in
the red shaded area in the top right corner, given by $f(x,y) = x + 2y
\geq z=3$? After approximately~$1.2 \cdot 10^7$
simulations, 100 such trajectories are found, with some of them
shown in light orange in
figure~\ref{fig:2mb-paths}. These can be considered typical
realizations for this extreme outcome, and allow us to
estimate $P_F^\eps(z) \in \left[6.71 \cdot 10^{-6},9.97 \cdot
  10^{-6}\right]$ as a $95\%$ confidence interval. While in principle
the same approach could be applied to much more complicated stochastic
models, such as SPDEs arising in atmosphere or ocean dynamics, it
quickly becomes infeasible due to the cost of performing such a large
number of simulations.

Instead, we generalize the strategy outlined in the previous
section. For the derivation, we make the following, compared to the
finite-dimensional case stronger
assumptions for technical reasons. To formulate them, we
introduce the solution map
\begin{align}
F \colon L^2([0,T],\RR^n) \to \RR\,, \quad &F[\eta] = f(\phi(T)), \nonumber\\
&\text{for } \begin{cases}
\dot{\phi} = b(\phi) + \sigma \eta\,,\\
\phi(0) = x\,.
\end{cases} 
\end{align}
Then, we assume for all $z \in \RR$:
\begin{enumerate}[leftmargin=*]
\item There is a unique instanton on the $z$-levelset of $F$, $\eta_z
  \in F^{-1}(\{z\}) \subset L^2([0,T], \RR^n)$, that minimizes the
  function $\tfrac{1}{2} \norm{\cdot}^2_{L^2([0,T], \RR^n)}$. There
  exists a Lagrange multiplier $\lambda_z \in \RR$ with $\eta_z =
  \lambda_z \left.\fdv{F}{\eta}\right|_{\eta_z}$ as a first-order
  necessary condition.  We define the large deviation rate
  function for the observable $f$ as
\begin{align}
I_F \colon
\RR \to \RR\,, \quad I_F(z) := \tfrac{1}{2}
\norm{\eta_z}^2_{L^2([0,T], \RR^n)}\,.
\label{eq:obs-rate-func}
\end{align}
\item The map from observable value to minimizer $z \mapsto \eta_z$ is $C^1$. In particular $I_F'(z) =
\langle \eta_z, \dd \eta_z / \dd z \rangle_{L^2([0,T], \RR^n)} = \lambda_z
 \langle \left. \tfrac{\delta F}{\delta \eta} \right|_{\eta_z}, \tfrac{\dd \eta_z}{\dd z}  \rangle_{L^2([0,T], \RR^n)}
= \lambda_z$.
\item $\Id - \lambda_z \left.\nfdv{2}{F}{\eta}\right|_{
\eta_z}$ is positive definite.
\item The rate function $I_F$ is twice continuously differentiable and
  strictly convex, i.e.\ $I_F'' > 0$.
\end{enumerate}
Under these assumptions and using existing
theoretical results on precise Laplace asymptotics for small-noise
SDEs, in
appendix~\ref{sec:deriv}\ref{subsec:deriv-laplace-infinite-dim} we sketch a derivation of the following result:
For the extreme event
probability 
\begin{equation} \label{eq:tail-def}
P_F^\eps(z) = \PP {\Big[ F[\sqrt{\eps} \eta] \geq z \Big]} = \PP
{\Big[ f(X_T^\eps) \geq z \Big]}
\end{equation}
with $z > F(0)$,
the asymptotically sharp estimate~\eqref{eq:extreme-event-finite-dimensional-result}
holds in the same way as before. The leading order prefactor
is now given by
\begin{align}
C_F(z) = \left[2 I_F(z) \det \left(\Id - \lambda_z
\ppr_{\eta_z^\perp} \left. \nfdv{2}{F}{\eta} \right|_{
\eta_z} \ppr_{
\eta_z^\perp} \right) \right]^{-1/2}\,,
\label{eq:tail-prob-prefac-sde}
\end{align}
where $\det$ is now a Fredholm determinant, the second
variation $\delta^2 F / \delta \eta ^2$ of the solution map $F$ at $\eta = \eta_z$
is a linear trace-class operator on $L^2([0,T],\RR^n)$, and $\ppr$
denotes orthogonal projection in $L^2([0,T],\RR^n)$.

Applied to the model SDE~(\ref{eq:2d-example}), we must first compute
the optimal noise realization $\eta_z = \left(\eta_z(t) \right)_{t \in
  [0,T]}$, which has a corresponding optimal system trajectory $\phi_z
= \left( \phi_z(t) \right)_{t \in [0,T]}$.  This optimal trajectory,
shown blue dashed in figure~\ref{fig:2mb-paths}, describes the most
likely evolution of the SDE~(\ref{eq:2d-example}) from the initial
condition $(0,0)$ into the shaded region in the upper right corner,
thus leading to an event $f(X(T),Y(T))\ge z$. Second, through
equation~(\ref{eq:tail-prob-prefac-sde}), we can compute the prefactor
correction for this optimal noise realization. Inserted into
equation~(\ref{eq:extreme-event-finite-dimensional-result}), we
obtain $P_F^{\eps = 0.5}(z = 3) \approx 8.94 \cdot 10^{-6}$ as
an asymptotic, sampling-free estimate, which falls into the estimated
interval obtained with direct
sampling. The source code to reproduce all
results for this example is available in a public GitHub repository
\citep{Schorlepp-github}. 

We add some remarks on the setting:
\begin{enumerate}[leftmargin=*]
\item We focus on SDEs with additive noise~\eqref{eq:sde} for
simplicity. For
the more general case of ordinary It{\^o} SDEs with multiplicative noise
$\sigma = \sigma(X_t^\eps)$, the leading order prefactor can still
be computed explicitly, but involves a regularized Carleman-Fredholm
determinant $\det_2$ (see~\cite{simon:1977} for a definition) instead
of a Fredholm determinant $\det$,
because the second variation of $F$ is no longer guaranteed to be
trace-class~\citep{ben-arous:1988}.
The direct analogy to the finite-dimensional case
is only possible for additive noise.
\item We state the theoretical result and computational strategy
for ordinary stochastic differential equations, but will also apply
them numerically to SPDEs with
additive, spatially smooth Gaussian forcing. In this case, we expect a
direct generalization of the results for SDEs to hold.
\item Without any additional work, we also obtain a sharp estimate, in the sense of~\eqref{eq:sharp-def}, for the
probability density function $\rho_F^\eps$ of $f(X_T^\eps)$ at $z$ via
\begin{align}
\hspace{1cm}\rho_F^\eps(z) \overset{\eps \downarrow 0}{\sim} (2 \pi \eps)^{-1/2}
\lambda_z C_F(z)\,\exp\left\{-\frac1\eps I_F(z)\right\}\,.
\label{eq:pdf-estimate}
\end{align}
\end{enumerate}

From a practical point of view, the remaining question is how
to evaluate~\eqref{eq:obs-rate-func}
and~\eqref{eq:tail-prob-prefac-sde}, given a general
and possibly high-dimensional SDE~\eqref{eq:sde}.

\paragraph*{Main questions and paper outline.}
In the remainder of this paper, we will specifically answer the
following questions:
\begin{itemize}[leftmargin=*]
\item How to find the minimizer~$\eta_z$ to the differential equation
  constrained optimization problem~\eqref{eq:obs-rate-func}
  numerically? This question has been treated in detail in the
  literature for the setup at hand, and we give a brief summary of
  relevant references in section~\ref{subsec:first-order}.
\item How to evaluate the Fredholm determinant
  in~\eqref{eq:tail-prob-prefac-sde} numerically? We show in
  section~\ref{subsec:pref-fred-adj} how to use second-order adjoints
  to compute the application of the projected second variation
  operator
  \begin{align}
    A_z := \lambda_z \ppr_{
      \eta^\perp_z} \left. \nfdv{2}{F}{\eta} \right|_{\eta_z} \ppr_{
      \eta^\perp_z}
      \label{eq:az-def}
  \end{align}
  to functions (or, upon discretization, to vectors), which is the
  basis for iterative eigenvalue solvers. In
  section~\ref{subsec:pref-fred-adj-comp}, we discuss how this allows
  us to treat very large system dimensions $n$ as long as the rank
  of~$\sigma$ remains small.
\item How does this prefactor computation based on the
  dominant eigenvalues of the projected second variation operator
  theoretically relate to the alternative approach using symmetric
  matrix Riccati differential equations mentioned in the introduction?
  What are the advantages and disadvantages of the different approaches?
  We comment on these points in sections~\ref{subsec:dets-ricc}
  and~\ref{subsec:pref-fred-adj-comp}.
\item What is the probabilistic interpretation of the quantities
  encountered when evaluating~\eqref{eq:obs-rate-func}
  and~\eqref{eq:tail-prob-prefac-sde}? In how far can they be observed
  in direct Monte Carlo simulations of the SDE~\eqref{eq:sde}? This is
  the content of section~\ref{sec:prob-interp}.
\end{itemize}
After these theoretical sections, illustrated throughout via the
model SDE~\eqref{eq:2d-example}, we present two challenging examples
in section~\ref{sec:ex}: The probability of high waves in
the stochastic Korteweg--De Vries equation in section~\ref{subsec:kdv},
and the probability of high strain events in the stochastic
three-dimensional incompressible Navier--Stokes equations in
section~\ref{subsec:nse}. All technical derivations can be found in
Appendix~\ref{sec:deriv}.

\section{Numerical rate function and prefactor evaluation}
\label{sec:pref-theo}

In this section, we show how the instanton and prefactor for the
evaluation of the asymptotic tail probability
estimate~\eqref{eq:extreme-event-finite-dimensional-result}
can be computed in practice for a general, possibly high-dimensional
SDE~\eqref{eq:sde}, and illustrate the procedure for the
model SDE~\eqref{eq:2d-example}. Both finding the
instanton (section~\ref{subsec:first-order}) and the
prefactor~(section~\ref{subsec:pref-fred-adj})
require the solutions of differential
equations of a complexity comparable to the original SDE. They therefore
become realistic to evaluate numerically even for fairly large
problems, provided tailored methods are used, as summarized in
section~\ref{subsec:pref-fred-adj-comp}. Additionally, we compare the
adjoint-based Fredholm determinant computation to the approach based on
matrix Riccati differential equations
in sections~\ref{subsec:dets-ricc}
and~\ref{subsec:pref-fred-adj-comp}.

\subsection{First variations and finding the instanton}
\label{subsec:first-order}

Here, we discuss the differential equation-constrained
optimization problem
\begin{align}
\eta_z =  \argmin_{\begin{subarray}{c}\eta\in L^2([0,T],\RR^n)\\
  \text{s.t. }F[\eta] = z
    \end{subarray}}  \;
  \frac{1}{2}
\norm{\eta}_{L^2([0,T],\RR^n)}^2,
\label{eq:min-prob-eta}
\end{align}
that determines the instanton noise $\eta_z$,
and briefly review how it can be solved numerically. We reformulate the
first-order optimality condition
\begin{align}
\eta_z = \lambda_z \left. \fdv{F}{\eta} \right|_{\eta_z}
\label{eq:first-order-opt}
\end{align}
by evaluating the first variation using
an adjoint variable as reviewed by~\cite{plessix:2006, HinzePinnauUlbrichEtAl09}.
For any $\eta \in L^2([0,T],\RR^n)$, we find
$\fdv{(\lambda F)}{\eta} = \sigma^\top \theta$,
where the adjoint variable $\theta$ (also called conjugate momentum)
is found via solving
\begin{align}
\begin{cases}
\dot \phi = b(\phi) + \sigma \eta\,, \quad &\phi(0) = x\,,\\
\dot \theta = - \nabla b^\top(\phi) \theta\,, \quad &\theta(T)
= \lambda \nabla f(\phi(T))\,.
\end{cases}
\label{eq:first-order-adj-eq}
\end{align}
With $a = \sigma \sigma^\top$, we recover
from~\eqref{eq:first-order-opt} the well-known instanton equations,
formulated only in term of the state variable $\phi_z$ and
its adjoint variable $\theta_z$ with optimal noise $\eta_z
= \sigma^\top \theta_z$:
\begin{align}
\begin{cases}
\dot \phi_z = b(\phi_z) + a \theta_z\,, \quad &\phi_z(0) = x\,,
\quad f(\phi_z(T)) = z,\\
\dot \theta_z = - \nabla b^\top(\phi_z) \theta_z\,,
\quad &\theta_z(T) = \lambda_z \nabla f(\phi_z(T))\,.
\end{cases}
\label{eq:inst-eq}
\end{align}
The rate function is given by $I_F(z) = \tfrac{1}{2}
\left \langle \theta_z, a \theta_z \right \rangle_{L^2([0,T],\RR^n)}$.
When formulating the optimization problem in the state variable $\phi$ instead
of the noise $\eta$, the instanton equations~\eqref{eq:inst-eq} are directly
obtained as the first-order necessary condition for a minimizer of the
Freidlin-Wentzell~\citep{freidlin-wentzell:2012} action functional $S$ with
\begin{align}
S[\phi] = \frac{1}{2} \int_0^T \left \langle \dot{\phi} - b(\phi),
a^{-1} \left[\dot{\phi} - b(\phi) \right] \right \rangle_n \dd t\,.
\end{align}
The numerical minimization of this functional for both ordinary
and partial stochastic differential equations is discussed
e.g.\ by~\cite{e-ren-vanden-eijnden:2004,grafke-grauer-schaefer:2015,grafke-grauer-schindel:2015,grafke-vanden-eijnden:2019,schorlepp-grafke-may-etal:2022}. Conceptually,
the minimization problem~\eqref{eq:min-prob-eta} is a deterministic
distributed optimal control problem on a finite time horizon with a final
time constraint on the state
variable~\citep{lewis-vrabie-syrmos:2012,herzog-kunisch:2010}. The
final-time constraint can be eliminated e.g.\ using penalty methods.
Alternatively, for a convex rate function, a primal-dual
strategy~\citep{boyd-vandenberghe:2004} with minimization
of $\tfrac{1}{2} \norm{\cdot}^2 - \lambda F$ at fixed $\lambda$ can
be used. If estimates for a range of $z$ are desired, one can solve
the dual problem for various $\lambda$, which effectively computes the
Legendre-Fenchel transform $I_F^*(\lambda)$, and invert afterwards. If
the rate function is not convex, the observable~$f$ can be reparameterized
to make this possible~\citep{alqahtani-grafke:2021}. To solve the
unconstrained problems of the general form~$ \min
\tfrac{1}{2} \norm{\cdot}^2 - \lambda (F - z) + \tfrac{\mu}{2}(F - z)^2$,
gradient-based methods with an adjoint evaluation~\eqref{eq:first-order-adj-eq}
can be used, e.g.\
\cite{schorlepp-grafke-may-etal:2022} use an L-BFGS solver. \cite{simonnet:2022} used a deep learning approach instead.
For high-dimensional problems such as multi-dimensional fluids, it may be
necessary to use checkpointing for the gradient evaluation, and to
use $\rank \sigma \ll n$ if applicable to reduce memory
costs~\citep{grafke-grauer-schindel:2015}. We comment on this point in
more detail in section~\ref{subsec:pref-fred-adj-comp}. Using second order
adjoints as in the next section would also make it possible to implement
a Newton solver, cf.~\cite{hinze-kunisch:2001,hinze-walther-sternberg:2006,sternberg-hinze:2010,cioaca-alexe-sandu:2012}.

For the model SDE~\eqref{eq:2d-example}, the instanton
equations~\eqref{eq:inst-eq} read
\begin{align}
&\begin{cases}
\dv{}{t} \left(\begin{array}{c}
\phi_1\\\phi_2
\end{array} \right) = \left(\begin{array}{c}
-\phi_1\\-4\phi_2
\end{array} \right) + \left(\begin{array}{c}
-\phi_1 \phi_2\\\phi_1^2
\end{array} \right)+ \left(\begin{array}{c}
\theta_1\\\tfrac{1}{4}\theta_2
\end{array} \right)\,,\\
\dv{}{t} \left(\begin{array}{c}
\theta_1\\\theta_2
\end{array} \right) = \left(\begin{array}{c}
+\theta_1\\+4\theta_2
\end{array} \right) + \left(\begin{array}{c}
\phi_2 \theta_1 - 2 \phi_1 \theta_2\\\phi_1\theta_1
\end{array} \right)\,,
\end{cases}\nonumber\\
\text{with }&\begin{cases}
\left(\begin{array}{c}
\phi_1(0)\\\phi_2(0)
\end{array} \right) = \left(\begin{array}{c}
0\\0
\end{array} \right)\,,
\quad \phi_1(T) + 2 \phi_2(T) = z,\\
\left(\begin{array}{c}
\theta_1(T)\\\theta_2(T)
\end{array} \right) = \lambda_z \left(\begin{array}{c}
1\\2
\end{array} \right)\,.
\end{cases}
\label{eq:inst-eq-2d-example}
\end{align}
We implemented a simple gradient
descent (preconditioned with $a^{-1}$) using adjoint evaluations of the
gradient and an Armijo line search (available in
the GitHub repository \citep{Schorlepp-github}) to find the instanton
for the model SDE~\eqref{eq:2d-example}. The state equation is
discretized using explicit Euler steps with an integrating factor, and the gradient
is computed exactly on a discrete level, i.e.\ ``discretize,
then optimize''. To find the instanton for a given $z$, we use
the augmented Lagrangian method. For each subproblem at fixed Lagrange
multiplier $\lambda$ and penalty parameter $\mu$, gradient descent is
performed until the gradient norm has been reduced by a given factor
compared to its initial value. All of these aspects are summarized in more
detail by~\cite{schorlepp-grafke-may-etal:2022}. 
The resulting optimal state variable trajectory $\phi_z$ for $z = 3$ for the
model SDE~\eqref{eq:2d-example} is shown in figure~\ref{fig:2mb-paths}.

\subsection{Second variations and prefactor computation via dominant eigenvalues}
\label{subsec:pref-fred-adj}

Similarly to the previous section, the second variation is also readily
evaluated in the adjoint formalism. With this prerequisite, we are able
to use iterative eigenvalue solvers to approximate the
Fredholm determinant $\det(\Id - A_z)$.
For a comprehensive introduction to the numerical computation of
Fredholm determinants, as well as theoretical results on approximate
evaluations using integral quadratures, see~\cite{bornemann:2010}.
However, in contrast to~\cite{bornemann:2010}, we deal with possibly spatially
high-dimensional problems, such as the example in
section~\ref{subsec:nse}. Hence, we use iterative algorithms to compute
the dominant eigenvalues to keep the number of operator evaluations
manageable.

Another application of the adjoint state method shows that
applying the second functional derivative of the solution
map $F$ at $\eta\colon[0,T] \to \RR^n$ to a fluctuation $\delta
\eta\colon[0,T] \to \RR^n$ results in
$\nfdv{2}{(\lambda F)}{\eta} \delta \eta = \sigma^\top \zeta$,
where $\zeta$ is found via solving
\begin{align}
&\begin{cases}
\dot \gamma =\nabla b(\phi) \gamma + \sigma \delta \eta\,,\\
\dot \zeta = -\left \langle \nabla^2 b(\phi), \theta \right
\rangle_n \gamma - \nabla b^\top(\phi) \zeta\,,
\end{cases} \nonumber\\
\text{with }&\begin{cases}
\gamma(0) = 0\,,\\
\zeta(T)
= \lambda \nabla^2 f(\phi(T)) \gamma(T)\,.
\end{cases}
\label{eq:second-order-adj-eq}
\end{align}
Here, we use the short-hand notation $\left[ \left \langle
\nabla^2 b(\phi), \theta \right \rangle_n\right]_{ij} =
\sum_{k=1}^n \partial_i \partial_j b_k(\phi) \theta_k$. The
trajectories $\phi$ and $\theta$ in \eqref{eq:second-order-adj-eq} are determined
via~\eqref{eq:first-order-adj-eq} from~$\eta$. Note that the second order
equations~\eqref{eq:second-order-adj-eq} are simply
the linearization of~\eqref{eq:first-order-adj-eq}.
Together with
the projection operator $\ppr_{\eta^\perp_z}$ acting as
\begin{align}
(\ppr_{\eta^\perp_z} \delta \eta)(t) = \delta \eta(t) -
\frac{\left \langle \eta_z, \delta \eta \right \rangle_{L^2([0,T],
\RR^n)}}{\norm{\eta_z}^2_{L^2([0,T],\RR^n)}} \eta_z(t)
\end{align}
for $t \in [0,T]$, we are now in a position to evaluate the
application of the operator $A_z$, as defined in~\eqref{eq:az-def},
to any function
$\delta \eta \colon [0,T] \to \RR^n$.
Denoting the eigenvalues of the trace-class operator $A_z$
by $\mu_z^{(i)} \in (-\infty, 1)$, the Fredholm determinant in
the prefactor~\eqref{eq:tail-prob-prefac-sde} is given by
$\det(\Id - A_z) = \prod_{i=1}^\infty (1 - \mu_z^{(i)})$,
with $\abs{\mu_z^{(i)}} \xrightarrow{i \to \infty} 0$ in such a way that
the product converges. An iterative eigenvalue solver relying
solely on matrix-vector multiplication, thus avoiding the explicit
storage of the possibly large discretized operator $A_z$ as
an $(n_t \cdot n)\times (n_t \cdot n)$ matrix, can now be used numerically
to find a finite number of dominant eigenvalues of $A_z$
with absolute value larger than some thresholds, and
approximate $\det(\Id - A_z)$ using these.

\begin{figure}
\centering
\includegraphics[width = .48 \textwidth]{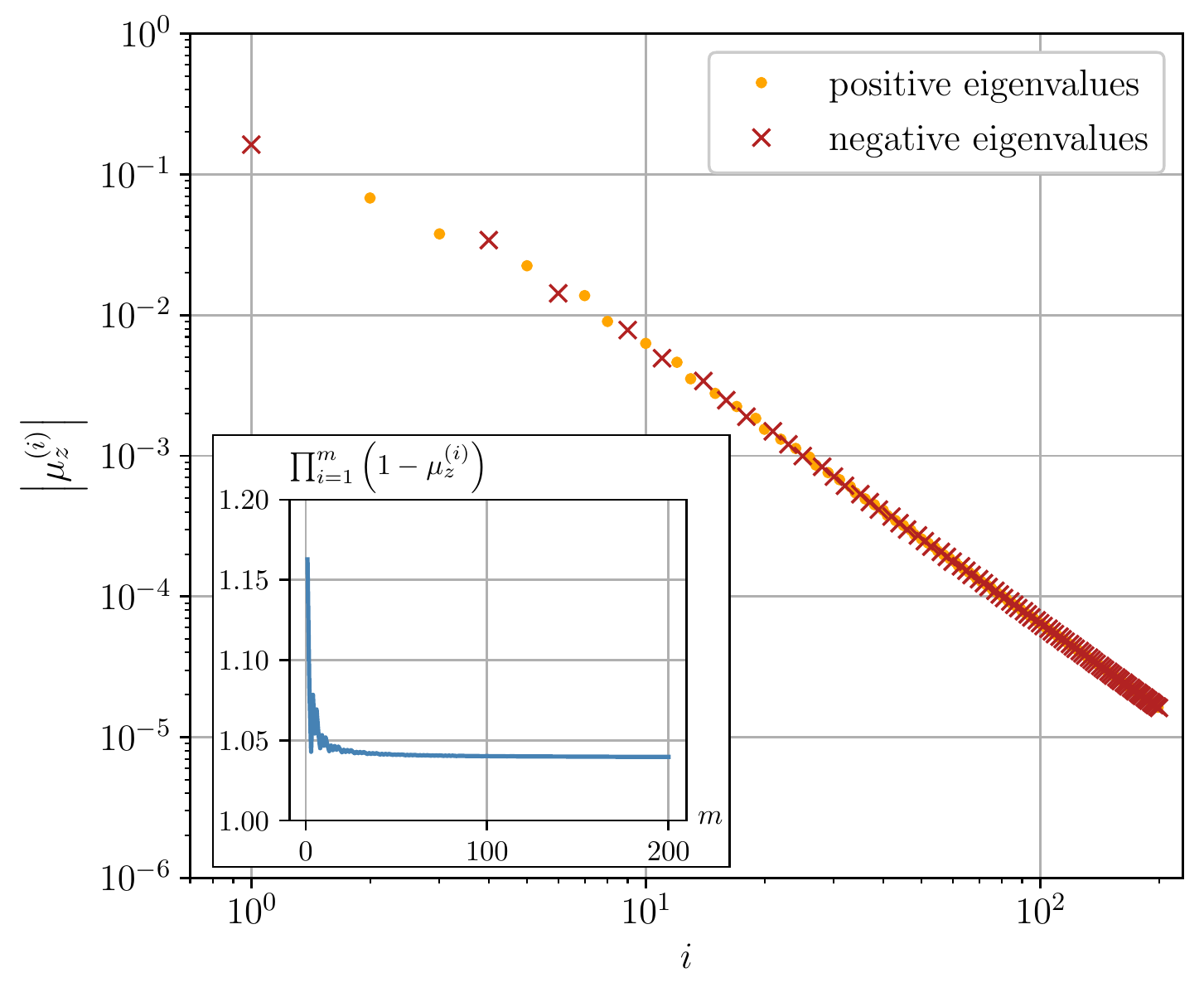}
\caption{Result of numerically computing 200 eigenvalues $\mu_z^{(i)}$ with largest
absolute value of $A_z$ for the example SDE~\eqref{eq:2d-example} with
$z = 3$. Discretization of~\eqref{eq:second-order-adj-eq-2d-examp} was done with step size $\Delta t = 5 \cdot 10^{-4}$, hence the dimension of the discretized path space variables is $4000$ here. Main figure: absolute value of
the eigenvalues~$\mu_z^{(i)}$. Inset: Finite product $\prod_{i = 1}^m
\left(1 - \mu_z^{(i)} \right)$ for different $m$ as an approximation for
the Fredholm determinant $\det (\Id - A_z)$. We see that the
eigenvalues rapidly decay to zero in this example, that similarly, the
cumulative product in the inset quickly converges, and that the
final estimate $\det (\Id - A_z) \approx \prod_{i = 1}^{200}
\left(1 - \mu_z^{(i)} \right) \approx 1.0397$ is in fact close
to $1$ in this example.}
\label{fig:2mb-evals}
\end{figure}

For the model example SDE~\eqref{eq:2d-example},
linearizing the state and first order adjoint
equations~\eqref{eq:first-order-adj-eq}, the second
order adjoint equations for~\eqref{eq:2d-example} become
\begin{align}
&\begin{cases}
\dv{}{t} \left(\begin{array}{c}
\gamma_1\\\gamma_2
\end{array} \right) = -\left(\begin{array}{c}\gamma_1\\4 \gamma_2\end{array}\right) + \left( \begin{array}{c}
- \gamma_1 \phi_2 - \phi_1 \gamma_2\\2\phi_1 \gamma_1
\end{array} \right) + \left(\begin{array}{c}
\delta \eta_1\\ \tfrac{1}{2}\delta \eta_2
\end{array} \right)\,,\\
\dv{}{t} \left(\begin{array}{c}
\zeta_1\\\zeta_2
\end{array} \right) = \left(\begin{array}{c}
\zeta_1\\4 \zeta_2 \end{array}\right) + \left(\begin{array}{c}
 \gamma_2 \theta_1+\phi_2 \zeta_1
-2 \gamma_1 \theta_2 - 2 \phi_1 \zeta_2 \\ \gamma_1 \theta_1 + \phi_1 \zeta_1
\end{array} \right) \,,
\end{cases} \nonumber\\
\text{with }&\begin{cases}
\left(\begin{array}{c}
\gamma_1(0)\\\gamma_2(0)
\end{array} \right) = \left(\begin{array}{c}
0\\0
\end{array} \right)\,,\\
\left(\begin{array}{c}
\zeta_1(T)\\\zeta_2(T)
\end{array} \right) = \left(\begin{array}{c}
0\\0
\end{array} \right)\,.
\end{cases}
\label{eq:second-order-adj-eq-2d-examp}
\end{align}
We implemented a simple Euler solver for these equations for a given
discretized input vector $\delta \eta \in \RR^{2 (n_t + 1)}$ in the
python code \citep{Schorlepp-github} as a
subclass of scipy.sparse.linalg.LinearOperator.  To set up this
operator, we supply the instanton data $(\phi_z, \theta_z, \lambda_z)
\in \RR^{2 (n_t + 1)} \times \RR^{2 (n_t + 1)} \times \RR$ as found
using the methods of the previous section~\ref{subsec:first-order}.
The LinearOperator class, for which we only need to supply a matrix
vector multiplication method instead of having to store the full
matrix $\in \RR^{2 (n_t + 1) \times 2 (n_t + 1)}$, can then be used
with any iterative eigenvalue solver. Here, we use the implicitly
restarted Arnoldi method of ARPACK~\citep{lehoucq-sorensen-yang:1998},
wrapped as scipy.sparse.linalg.eigs in python. Note that in this
example, storing the full matrix would be feasible, and the Riccati
method discussed in the next section is faster to compute the
prefactor.  However, we are interested in a scalable approach for
large $n$, where, as discussed in
section~\ref{subsec:pref-fred-adj-comp} and shown in
section~\ref{sec:ex}, the Riccati approach becomes infeasible. We show
the results of computing 200 eigenvalues with largest absolute value
of the projected second variation operator $A_z$ for $z=3$ in
figure~\ref{fig:2mb-evals}.

\subsection{Alternative: prefactor computation via matrix Riccati differential equations}
\label{subsec:dets-ricc}
In Appendix~\ref{sec:deriv}\ref{subsec:deriv-fred-zeta}, we
motivate via formal manipulations that the
prefactor~\eqref{eq:tail-prob-prefac-sde} can also be expressed via
the following ratio
of zeta-regularized functional determinants~\citep{ray-singer:1971}
of second order differential operators,
instead of a Fredholm determinant of an integral operator. This
prefactor expression is more natural from the statistical physics
point of view, where path integrals in the field variable $\phi$ instead of
the noise $\eta$ are typically considered,
cf.~\cite{zinn-justin:2021}. We obtain
\begin{align}
C_F(z) =& \sqrt{I_F''(z)}\lambda_z^{-1} \left(\frac{\Det_{{\cal A}_{\lambda_z}}
  \left(\Omega[\phi_z] \right)}
  {\Det_{{\cal A}_0} \left(\Omega[\phi_0] \right)}
  \right)^{-1/2} \times \nonumber\\ &\quad\times \exp\left\{-\tfrac12 \int_0^T \left(\nabla\cdot
  b(\phi_z) - \nabla\cdot b(\phi_0)\right)\,\dd t\right\}\,,
  \label{eq:det-ratio}
\end{align}
in accordance with~\cite{schorlepp-grafke-grauer:2023},
where it was derived directly through path integral computations.
Here, $\Omega$ is the Jacobi operator of the Freidlin-Wentzell action functional
as defined in the appendix~\ref{sec:deriv}\ref{subsec:deriv-fred-zeta}, and
the subscript of the zeta-regularized determinants $\Det$ denotes the
boundary conditions under which the
determinants of the differential operators are computed.
Naively evaluating the determinant ratio
in~\eqref{eq:det-ratio} by numerically finding the eigenvalues of the appearing
differential operators is typically not feasible. This is due to the fact that
both operators posses unbounded spectra with the same asymptotic behavior of the
eigenvalues, which requires computing the \textit{smallest} eigenvalues of both
operators. A threshold for this computation is difficult to set, and while the
eigenvalues of both operators should converge to each other as they increase,
numerical inaccuracies tend to increase for the larger eigenvalues.
Fortunately, there exists theoretical results regarding the
computation of such
determinant ratios exactly and in a closed form by solving initial value
problems~\citep{gelfand-yaglom:1960,levit-smilansky:1977,forman:1987,kirsten-mckane:2003}. 
Using the results of \citet{forman:1987}, the
prefactor~\eqref{eq:det-ratio} can be computed by solving the symmetric
matrix Riccati differential equation
\begin{align}
\begin{cases}
\dot{Q}_z = a + Q_z \nabla b\left(\phi_z
\right)^\top +
 \nabla b\left(\phi_z \right) Q_z + Q_z \left \langle \nabla^2
b(\phi_z), \theta_z\right \rangle_n Q_z \,,\\
Q_z(0) = 0_{n \times n} \in \RR^{n \times n}\,.
\end{cases}
\label{eq:riccati-fw}
\end{align}
for $Q_z \colon [0,T] \to \RR^{n \times n}$ and then evaluating
\begin{align}
C_F(z) &=
\lambda_z^{-1} \exp\left\{\frac{1}{2} \int_0^T
\trace \left[\left \langle \nabla^2 b(\phi_z), \theta_z \right
\rangle_n  Q_z \right] \dd t\right\} \times \nonumber\\
&\times \left[{\det} \left( U_z \right) \left \langle \nabla f(\phi_z(T)),
Q_z(T) U_z^{-1} \nabla f(\phi_z(T)) \right \rangle_n  \right]^{-1/2}
\label{eq:cf-ric}
\end{align}
with
\begin{align}
U_z := 1_{n \times n} - \lambda_z \nabla^2 f
\left(\phi_z(T) \right) Q_z(T) \in \RR^{n \times n}\,.
\end{align}
This result in terms of a Riccati matrix differential equation
is also natural from a stochastic
analysis perspective (WKB analysis of the Kolmogorov backward
equation~\citep{grafke-schaefer-vanden-eijnden:2021}), or a time-discretization
of the path integral perspective (recursive evaluation
method~\citep{schorlepp-grafke-grauer:2021}). To give intuition for the
Riccati differential equation~\eqref{eq:riccati-fw},
note that by letting $Q_z = \gamma \zeta^{-1}$
with $\gamma(0) = 0_{n \times n}$ and $\zeta(0) = 1_{n \times n}$,
the approach amounts to solving
\begin{align}
\begin{cases}
\dot \gamma =\nabla b(\phi) \gamma + a \zeta\,,
\quad &\gamma(0) = 0_{n \times n}\,,\\
\dot \zeta = -\left \langle \nabla^2 b(\phi), \theta \right
\rangle_n \gamma - \nabla b^\top(\phi) \zeta\,, \quad &\zeta(0)
= 1_{n \times n},
\end{cases}
\label{eq:riccati-trafo}
\end{align}
as an initial value problem, whereas the eigenvalue
problem $\nfdv{2}{(\lambda F)}{\eta} \delta \eta = \mu \delta \eta$ corresponds
to the boundary value problem
\begin{align}
&\begin{cases}
\dot \gamma =\nabla b(\phi) \gamma + \mu^{-1} a \zeta\,,\\
\dot \zeta = -\left \langle \nabla^2 b(\phi), \theta \right
\rangle_n \gamma - \nabla b^\top(\phi) \zeta\,,
\end{cases} \nonumber\\
\text{with }&\begin{cases}
\gamma(0) = 0\,,\\
\zeta(T)
= \lambda \nabla^2 f(\phi(T)) \gamma(T)\,.
\end{cases}
\end{align}
This means that to evaluate the functional determinant prefactor
via the Riccati approach, we consider functions in the kernel of the operator $\Id - \lambda \delta^2 F / \delta \eta^2$,
i.e.\ eigenfunctions belonging to the eigenvalue $0$,
but under modified boundary conditions of the operator. In practice,
instead of finding the dominant eigenvalues of the
integral operator $A_z$ of section~\ref{subsec:pref-fred-adj} that
acts on functions $\delta \eta \colon [0,T] \to \RR^n$, we can integrate 
a single matrix-valued initial value problem for $Q_z \colon [0,T]
\to \RR^{n \times n}$ as presented in this section.
Even though the Riccati equation~\eqref{eq:riccati-fw}, in contrast to the
linear system~\eqref{eq:riccati-trafo}, is a nonlinear differential equation,
it is nevertheless advisable to solve~\eqref{eq:riccati-fw} instead
of~\eqref{eq:riccati-trafo} numerically because the equation for
$\zeta$ in~\eqref{eq:riccati-trafo} has to be integrated in the unstable
time direction for the right-hand side term $-\nabla b(\phi)^\top \zeta$. 
Note also that, depending on the system and observable at hand, the solution
of the Riccati equation~\eqref{eq:riccati-fw} may pass through removable
singularities in $(0,T)$ whenever $\zeta(t)$ in~\eqref{eq:riccati-trafo} becomes
non-invertible, hence direct numerical integration of~\eqref{eq:riccati-fw}
may require some care (see \cite{schiff-snider:1999} and references therein).

For the two-dimensional model SDE~\eqref{eq:2d-example}, the forward Riccati
equation for the symmetric matrix $Q = Q_z \colon [0,T] \to
\RR^{2 \times 2}$ along $(\phi, \theta) = (\phi_z, \theta_z)$ becomes
\begin{align}
&\dv{}{t}\left(\begin{array}{cc}
Q_{11} & Q_{12}\\
Q_{12} & Q_{22}
\end{array} \right) = \left(\begin{array}{cc}
1 & 0\\
0 & \tfrac{1}{4}
\end{array} \right) - \left(\begin{array}{cc}
2 Q_{11} & 5 Q_{12}\\
5 Q_{12} & 8 Q_{22}
\end{array} \right) \nonumber\\
&+ \left[ \left(\begin{array}{cc}
-\phi_2 & -\phi_1\\
2 \phi_1 & 0
\end{array} \right) \left(\begin{array}{cc}
Q_{11} & Q_{12}\\
Q_{12} & Q_{22}
\end{array} \right) \right] + [\dots]^\top \nonumber\\
&+\left(\begin{array}{cc}
Q_{11} & Q_{12}\\
Q_{12} & Q_{22}
\end{array} \right) \left(\begin{array}{cc}
2 \theta_2 & - \theta_1\\
- \theta_1 & 0
\end{array} \right) \left(\begin{array}{cc}
Q_{11} & Q_{12}\\
Q_{12} & Q_{22}
\end{array} \right)\,,
\end{align}
where $[\dots]$ stands for a repetition of the preceding term.
We solve the Riccati equation with Euler steps with integrating factor
in ~\citep{Schorlepp-github}, and
use it to evaluate~\eqref{eq:cf-ric}. We do not encounter any numerical
problems or singularities in this example. The result for $C_F(z = 3)$ agrees
with the Fredholm determinant computation using dominant eigenvalues in the
previous section~\ref{subsec:pref-fred-adj}.

\subsection{Computational efficiency considerations}
\label{subsec:pref-fred-adj-comp}

In this section, we compare the two prefactor computation methods of
sections~\ref{subsec:pref-fred-adj} and~\ref{subsec:dets-ricc} using either
dominant eigenvalues of the trace-class operator $A_z$ evaluated
via~\eqref{eq:second-order-adj-eq}, or the Riccati matrix differential
equation~\eqref{eq:riccati-fw}, in terms of their practical applicability
as well as computational and memory cost for large system dimensions $n \gg 1$.

For the eigenvalue-based approach, we know that $\prod_{i = 1}^m
\left(1 - \mu_z^{(i)} \right) \xrightarrow{m \to \infty} \det(\Id - A_z)$
converges in theory, but it is difficult to give bounds on the required
number of eigenvalues for an approximation of the Fredholm determinant to a
given accuracy. In all examples considered
in this paper, at most a few 100 eigenvalues turned out to be necessary
for accurate results, even for the three-dimensional Navier--Stokes
equations in section~\ref{subsec:nse} as a high-dimensional ($n = 3 \cdot 128^3 \approx 6.3 \cdot 10^6$) and strongly
nonlinear example. The number of dominant eigenvalues of $A_z$ to be computed
to achieve a desired accuracy is robust with respect to
the temporal resolution
and only depends on the (effective, see below) dimension of the system
and the level of nonlinearity in the system.
In any case, to obtain~$m$ eigenvalues of~$A_z$ with largest
absolute value, iterative eigenvalue
solvers, either using Krylov subspace methods or randomized algorithms, typically require a number of evaluations of the operator that is equal
to a constant times $m$~\citep{halko-martinsson-tropp:2011}.
Each evaluation of~$A_z$ consists of solving two ODEs or
PDEs~\eqref{eq:second-order-adj-eq} with comparable computational complexity
to the original SDE. We comment on memory requirements below.

Compared to this, the Riccati approach requires the numerical solution
of a single $n \times n$ symmetric matrix differential equation as an
initial value problem. If $n$ is small, then this is clearly more
efficient than computing $m > n / 2$ eigenvalues.  However, there may
also be problems with the Riccati approach: On the one hand, this
approach requires a strictly convex rate function with $I_F''(z)> 0$
at~$z$, as can be seen from~\eqref{eq:det-ratio}. If this is not
satisfied, then a suitable convexification via reparameterization
needs to be carried out on a case-by-case
basis~\citep{alqahtani-grafke:2021}. While we assumed that the rate
function is convex to derive the
prefactor~\eqref{eq:tail-prob-prefac-sde} in terms of the Fredholm
determinant, this assumption is actually not necessary and the
eigenvalue-based approach remains feasible regardless of the convexity
of the observable rate function $I_F$. Finally, the eigenvalue
approach is easier to interpret, while it is not always immediately
clear why the Riccati solution may diverge (removable singularities
that can be remedied by a suitable choice of integration scheme versus
true singularities due to unstable or flat directions of the second
variation at the instanton).

We turn to the memory requirements of the prefactor computations strategies,
and in particular to their scaling with the system dimension $n$. Informally,
one can think of the Riccati matrix as defined in the (squared) state space
of the SDE, in contrast to the eigenvectors of $A_z$ that are defined in the noise
space that is potentially lower-dimensional. The Riccati
equation then integrates a dense $n \times n$ array in time by
performing $n_t$ consecutive times steps of~\eqref{eq:riccati-fw}
and evaluating~\eqref{eq:cf-ric} along the way. This is difficult to achieve
directly as soon as (semi-discretizations of) multi-dimensional SPDEs are
considered, which are relevant e.g.\ for realistic fluid or climate models.
Usually, large Riccati matrix differential equations, which also arise e.g.\ in
linear-quadratic regulator problems, are solved within some problem-specific
low-rank format, see e.g.~\cite{stillfjord:2018}. In contrast to this, the
vectors on which iterative eigenvalue solvers for the Fredholm-determinant
based approach need to operate are in general vectors of size $n_t \times n$.

\begin{figure*}
\centering
\includegraphics[width = \textwidth]{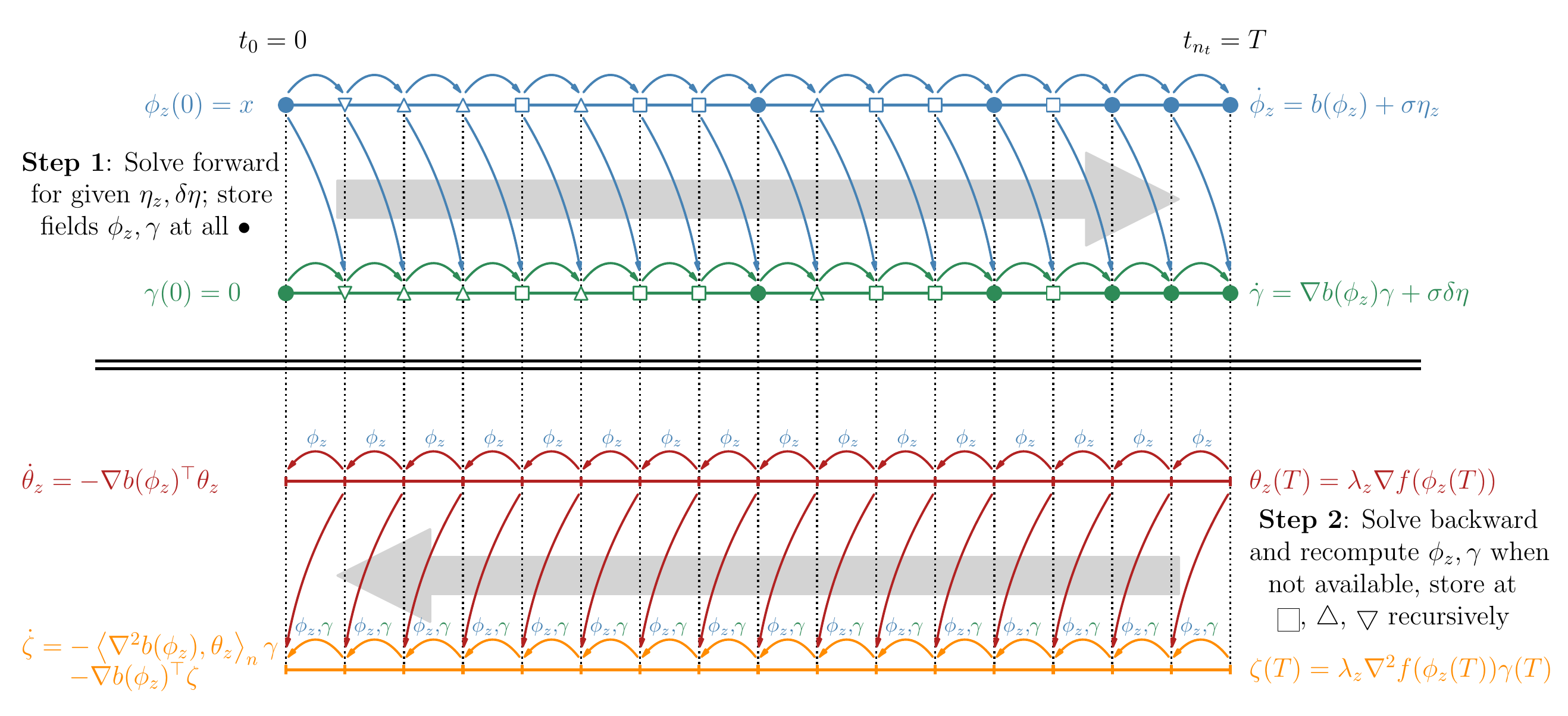}
\caption{Sketch of the checkpointing procedure used
to evaluate the second variation operator $\delta^2(\lambda F) / \delta \eta^2$
at $\eta_z$, applied to $\delta \eta$, for large system
dimensions $n \gg 1$ in a memory-efficient way. The instanton
noise $\eta_z$, the input noise fluctuation $\delta \eta$, and the
return vector $\sigma^\top \zeta$ are all stored as
dense $(n_t + 1, \rank \sigma)$-arrays for $\rank \sigma \ll n$. Given the
instanton noise~$\eta_z$ and a noise fluctuation~$\delta \eta$, the first step
consists of solving the state equation and linearized state equation
for $\phi_z$ and $\gamma$ simultaneously forward in time
from $t = 0$ to $t = T$, and storing the
fields $\phi_z(t_i) \in \RR^n$ and $\gamma(t_i) \in \RR^n$
only at the logarithmically spaced instances $t_i = \bullet$. Afterwards,
the first and second order adjoint
equations for~$\theta_z$ and~$\zeta$ are simultaneously solved backwards in
time from $t = T$ to $t = 0$ and $\sigma^\top \zeta(t_i)$ is
stored for each~$t_i$. Whenever $\phi_z(t_j)$ and $\gamma(t_j)$ are needed for
the time integration, but not available in storage already, the two forward equations are solved
again from the nearest preceding point in time when they are available,
and recursively stored at intermediate
steps
$\square$, $\bigtriangleup$,
$\bigtriangledown$, \dots All fields $\phi_z(t_i) \in \RR^n$
and $\gamma(t_i) \in \RR^n$ that are no longer needed during the
backwards integration are deleted from memory.}
\label{fig:checkp}
\end{figure*}

As an important class of examples, we now consider systems with large
spatial dimension $n \gg 1$, for which, however, only a few degrees of
freedom are  forced, such that the diffusion
matrix $a = \sigma \sigma^\top$ is singular and $\rank \sigma \ll n$.
Examples for this include fluid and turbulence
models with energy injection only on a compactly supported set of either
high or low spatial Fourier modes, or climate models with a limited number
of random parameters in the model~\citep{margazoglou-grafke-laio-etal:2021}.
In this case, it is straightforward to exploit the small rank of $\sigma$ 
within the eigenvalue-based approach to decrease the memory requirements
and apply the method even to very high-dimensional models, which we demonstrate
for the randomly forced three-dimensional Navier--Stokes equations
in section~\ref{subsec:nse} in this paper. The idea is that for the
eigenvectors $\delta \eta$ of $A_z$, clearly only $\rank \sigma$ many entries
are relevant due to the composition with $\sigma$ and $\sigma^\top$.
Eigenvalue solvers hence act on $n_t \times \rank \sigma$ vectors, which should
fit into memory. This is similar to the computation of the instanton itself,
where only the instanton noise $\eta_z$ as a $n_t \times \rank \sigma$ vector
is computed and stored explicitly, as discussed
by~\cite{grafke-grauer-schindel:2015,schorlepp-grafke-may-etal:2022}. The
remaining challenge is then to evaluate $A_z \delta \eta$ for
given $\delta \eta \in \RR^{n_t \times \rank \sigma}$ by solving the
second order adjoint equations~\eqref{eq:second-order-adj-eq},
\textit{without} storing the full, prohibitively large $n_t \times n$
arrays needed for $\phi_z$, $\gamma$ and $\theta_z$. Similar to the gradient
itself, evaluated via the first order adjoint
approach~\eqref{eq:first-order-adj-eq}, this is possible through
(static) checkpointing \citep{GriewankWalther00}, as illustrated in figure~\ref{fig:checkp}.
At the cost of having to integrate the first order adjoint equations
repeatedly for each noise vector $\delta \eta$ to which $A_z$ is applied,
and to recursively solve the forward equations for $\phi_z$ and $\gamma$
again and again, the memory requirements for the spatially dense fields
are only ${\cal O}\left(\log n_t \cdot n \right)$ this way.
The same problem is encountered and solved similarly in implementations
of Newton solvers for high-dimensional PDE-constrained optimal control
problems~\citep{hinze-kunisch:2001,hinze-walther-sternberg:2006,sternberg-hinze:2010,cioaca-alexe-sandu:2012}.
All in all, in contrast to the Riccati formalism, this permits an easy
and controlled strategy that enables to treat very large spatial
dimensions within the Fredholm-based prefactor approach, as long as
the diffusion matrix possesses a comparably small rank.
Note, however, that it is
still necessary that the number of eigenvalues needed to
approximate $\det(\Id - A_z)$ remains small for this approach to be
applicable in practice. We show numerically in section~\ref{subsec:nse}
that this is indeed the case for the three-dimensional Navier--Stokes equations
as an example. The discussion of this paragraph, with all relevant
scalings of computational and memory costs for the different approaches,
is briefly summarized in table~\ref{tab:cost-overview}.

\begin{table}
  \caption{Overview of computational and memory costs for finding the
    prefactor $C_F(z)$ either through solving the Riccati
    equation~\eqref{eq:riccati-fw}, or through determining $m$
    dominant eigenvalues of $A_z$. The system's spatial dimension is
    denoted by $n \gg 1$ and the noise correlation has $\rank \sigma
    \ll n$.  In the table, $c$ denotes the computational costs of
    integrating the original S(P)DE once from $t = 0$ to $t = T$. Any
    multiplicative constants were omitted for the costs listed in the
    table, the eigenvalue-based approach is assumed to use
    checkpointing as sketched in figure~\ref{fig:checkp}, the
    computational costs of evaluating the quadratic term
    in~\eqref{eq:riccati-fw} were ignored, and the complete instanton
    data is assumed to be known. The table shows that the
    eigenvalue-based approach indeed remains feasible for large $n$.}
\label{tab:cost-overview}
\begin{tabular}{l|c|c}
 & Riccati  & Eigenvalues \\
\hline
Memory costs & $n^2$ & $ n\cdot  \log n_t + n_t \cdot \rank \sigma$ \\
\hline
Computational costs & $n \cdot c$ & $m \cdot \log n_t \cdot c$ \\ \hline
\end{tabular}
\end{table}

In conclusion, we recommend using the Riccati equation only in sufficiently
``nice'' situations for small to moderate system dimensions $n$.
For such systems and diffusion matrices without low-rank properties, and
as long as no additional complications such as non-convex rate functions
or removable singularities of the Riccati solution are encountered,
it is faster than
the eigenvalue-based approach, and better suited to
analytical computations or approximations since it only involves the solution
of initial value problems, in contrast to the boundary value problems that
need to be solved to find eigenfunctions of the projected second
variation operator~$A_z$. On the other hand, the
Fredholm determinant computation through
dominant eigenvalues is easier to use and implement, requiring only
solvers for the original SDE, its adjoint, as well as their linearizations.
At the cost of introducing numerical errors and a step
size parameter $h>0$ that needs to be adjusted, one can also approximate
the second variation evaluations via
\begin{align}
\left.\nfdv{2}{\left(\lambda_z F \right)}{\eta}\right|_{\eta_z} \delta \eta
\approx \frac{1}{h} \left(\left.\fdv{\left(\lambda_z F \right)}{\eta}
\right|_{\eta_z + h \delta \eta} - \left.\fdv{\left(\lambda_z F \right)}{
\eta}\right|_{\eta_z} \right)
\end{align}
or other finite difference approximations, which does
not require implementing any second order variations. In this sense, both the
numerical instanton and leading-order prefactor computation can quickly be
achieved in a black-box like, non-intrusive way when solvers for the state
equation and its adjoint are available. Alternatively,
the adjoint solver, as well as solvers for the second order tangent and
adjoint equation can be obtained through
automatic differentiation~\citep{naumann:2011}. We also note that
in the context of the second order reliability method, there exist
further approximation methods that could be used here
for the Fredholm determinant prefactor,
e.g.\ by extracting information from the gradient based optimization
method that has been used to find the
instanton or design point~\citep{derkiureghian-stefano:1991}, or
through constructing a non-infinitesimal parabolic approximation
to the extreme event set~\citep{derkiureghian-lin-hwang:1987}.

In any case, for the scenario of possibly
multi-dimensional SPDEs with low-rank forcing, we argue that
the eigenvalue approach is to be preferred as
it leads to  natural approximations and a simpler implementation.
However, we remark that in the case of SDEs with multiplicative noise,
or SPDEs with spatially white noise that need to be renormalized such
as the Kardar--Parisi--Zhang (KPZ) equation, the Riccati
approach remains structurally unchanged~\citep{schorlepp-grafke-grauer:2023},
whereas the Fredholm determinant expression turns into a Carleman-Fredholm
determinant and an additional operator trace~\citep{ben-arous:1988}, which
could potentially be more costly to evaluate.

\section{Probabilistic interpretation via fluctuation covariances and transition tubes}
\label{sec:prob-interp}

In this section, we give an intuitive interpretation for some of the
quantities encountered in the previous sections. The second
variation quantifies the linearized dynamics of the SDE~\eqref{eq:sde} around
the most likely realization. This implies that dominating
eigenfunctions of the second variation correspond to fluctuation
modes that are most easily observable. Below, we confirm this with a
simple numerical experiment that relates the eigenfunction
information with the transition tube along a rare trajectory. The basic
object that we consider in this section
is the process $(X_t^\eps)_{t \in [0,T]}$ as $\eps \downarrow 0$,
conditioned on the rare outcome $f(X_T^\eps) = z$ at final time. In other
words, we consider only \textit{transition paths} between the fixed
initial state $x \in \RR^n$ and any final state in the target
set $f^{-1}(\{z\}) \subset \RR^n$. The path on which the transition path
ensemble concentrates as $\eps \downarrow 0$ is given by the state variable
instanton trajectory~$\phi_z$, i.e.\ the most likely way for the system to
achieve $f(X_T^\eps) = z$, since deviations from it are suppressed
exponentially~\citep{freidlin-wentzell:2012}.
One thus has
\begin{align}
\lim_{\eps \downarrow 0} \EE \left[ X_t^\eps \mid f(X_T^\eps)= z \right]
= \phi_z(t)
\end{align}
for the mean of the conditioned process. In this sense, by taking
conditional averages of direct Monte Carlo simulations
of~\eqref{eq:sde} as $\eps$ tends to 0, the instanton trajectory $\phi_z$
is directly observable, and the mean realization agrees with the most
likely one for $\eps \downarrow 0$.
This procedure is sometimes called filtering, and has been carried out
e.g.\ for the one-dimensional Burgers equation~\citep{grafke-grauer-schaefer:2013},
the three-dimensional Navier--Stokes
equations~\citep{grafke-grauer-schaefer:2015,schorlepp-grafke-may-etal:2022}
and the one-dimensional KPZ equation~\citep{hartmann-meerson-sasorov:2021}.
Using the results of the previous sections, we can, however, make this statement
more precise and state a central limit-type theorem for the conditioned
fluctuations at order $\sqrt{\eps}$ around the instanton: As $\eps \downarrow 0$,
the process $(X_t^\eps - \phi_z(t)) / \sqrt{\eps}$, conditioned on $f(X_T^\eps)
= z$, becomes centered Gaussian. It is hence fully characterized by its
covariance function
${\cal C}_z \colon [0,T] \times [0,T] \to \RR^{n \times n}$, given by
\begin{align}
{\cal C}_z(t,t') = \lim_{\eps \downarrow 0} \EE \left[\frac{(X_t^\eps -
\phi_z(t))\otimes(X_{t'}^\eps -
\phi_z(t'))}{\eps} \bigg\mid f(X_T^\eps) = z\right]\,.
\label{eq:cov-def}
\end{align}
We show in Appendix~\ref{sec:deriv}\ref{subsec:deriv-eig-cov} that
${\cal C}_z$ is fully determined through the 
ortho\textit{normal} eigenfunctions $\delta \eta^{(i)}_z$ of the projected
second variation operator $A_z$
with corresponding eigenvalues $\mu_z^{(i)}$ and associated state variable
fluctuations $\gamma^{(i)}_z$, the solution of the linearized state equation
\begin{align}
\dot{\gamma}^{(i)}_z = \nabla b (\phi_z) \gamma^{(i)}_z + \sigma
\delta \eta^{(i)}_z\,, \quad \gamma^{(i)}_z = 0\,,
\end{align}
via
\begin{align}
{\cal C}_z(t,t') = \sum_{i = 1}^\infty \frac{\gamma^{(i)}_z(t)
\otimes \gamma^{(i)}_z(t')}{1 - \mu_z^{(i)}}\,.
\label{eq:cov-eig}
\end{align}
In particular, computing the eigenvalues and eigenfunctions of $A_z$ yields
a complete characterization of the conditioned Gaussian fluctuations
around the instanton. As detailed in the example below, at small but
finite~$\eps$, ${\cal C}_z$ can be used to
approximate the distribution of transition paths at any time $t \in [0,T]$ as
multivariate normal ${\cal N}(\phi_z(t), \eps {\cal C}_z(t,t))$. Effectively,
in addition to the mean transition path at small noise, the instanton $\phi_z$,
we can also estimate the width and shape of the \textit{transition tube} around
it at any $t \in [0,T]$ without sampling within a Gaussian process
approximation of the conditioned SDE; see~\cite{vanden-eijnden:2006} for a
general introduction to transition path theory,
and~\cite{archambeau-cornford-oppper-etal:2007,lu-stuart-weber:2017} for
Gaussian process approximations of SDEs based on minimizing the path space
Kullback--Leibler divergence, which, in the small-noise limit and for
transition paths, reduce to the Gaussian process considered here.
Furthermore, one can show that the forward Riccati
approach of section~\ref{subsec:dets-ricc} recovers the \textit{final-time}
state variable fluctuation covariance via
\begin{align}
{\cal C}_z(T,T) &= Q_z(T) U_z^{-1} \nonumber\\
&\quad- \frac{\left(Q_z(T) U_z^{-1} \nabla f(\phi_z(T)
)\right)^{\otimes 2}}{\left \langle \nabla f(\phi_z(T)), Q_z(T)
U_z^{-1} \nabla f(\phi_z(T)) \right \rangle_n}\,.
\label{eq:cov-ric}
\end{align}
This directly follows by adapting the forward Feynman-Kac computation used
in remark 4 of~\cite{schorlepp-grafke-grauer:2021} to the present calculation
of the covariance function~\eqref{eq:cov-def} at final time $t = t' = T$.
Note that both, directly from~\eqref{eq:cov-ric}, as well as
from~\eqref{eq:cov-eig} after a short calculation, carried out in
Appendix~\ref{sec:deriv}\ref{subsec:deriv-bc}, one can see that these results
are consistent with the additional final time boundary condition for the
state variable fluctuations
\begin{align}
\lim_{\eps \downarrow 0} \left \langle \nabla f(\phi_z(T)),
\frac{X_T^\eps - \phi_z(T)}{\sqrt{\eps}} \right \rangle_n = 0,
\label{eq:final-bc}
\end{align}
almost surely, when conditioning on $f(X_T^\eps) = z$.
In words, the conditioned Gaussian fluctuations at final time are constrained
to the tangent plane of the equi-observable hypersurface $f^{-1}(\{z\})$
at the point $\phi_z(T)$.

\begin{figure*}
\centering
\includegraphics[width = \textwidth]{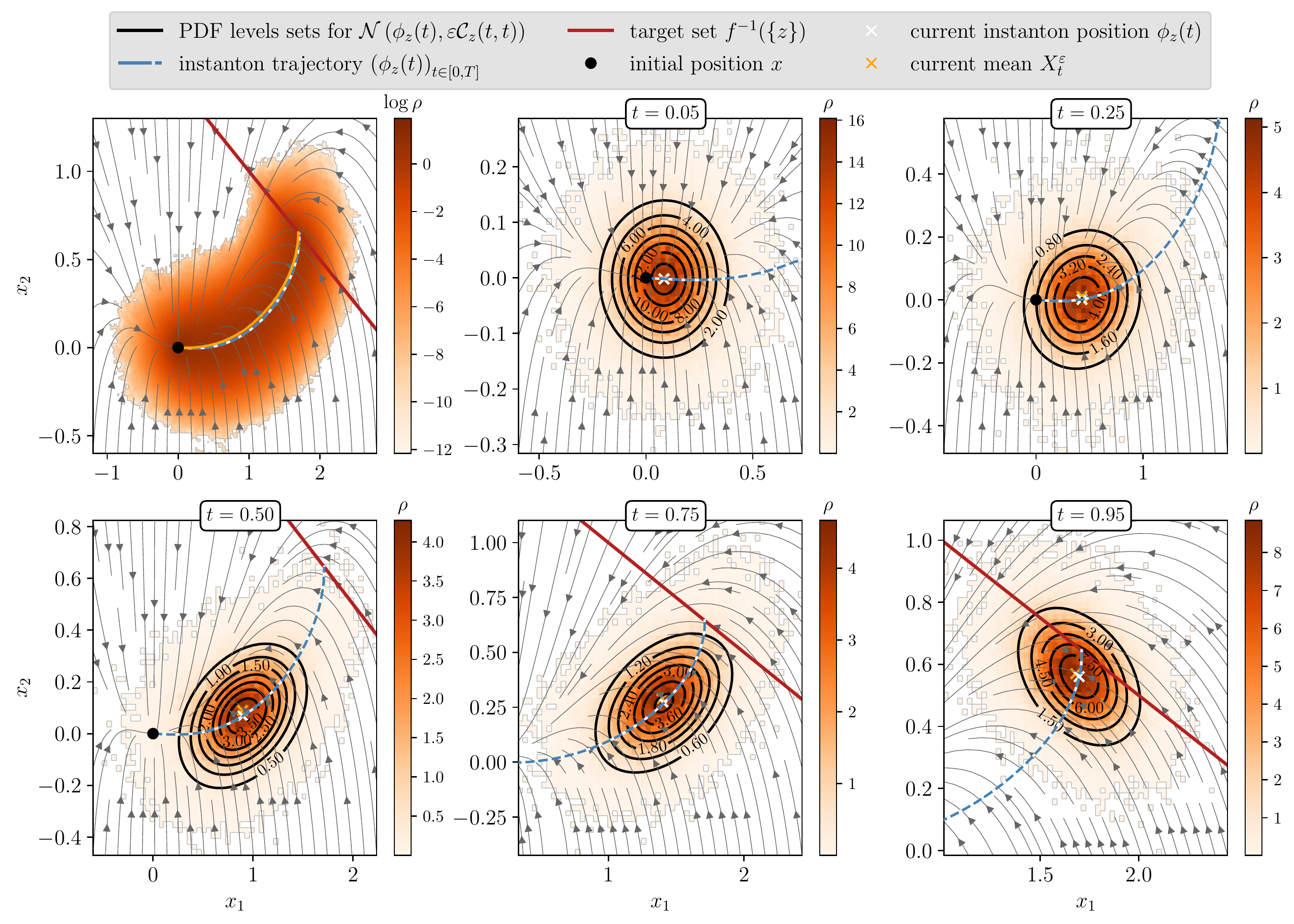}
\caption{Results of numerically computing $10^5$ transition paths
from $x = 0$ to the target set $f^{-1}(\{z\})$ for the
model SDE~\eqref{eq:2d-example}
with $z = 3$ and $\eps = 0.5$ using instanton-based importance
sampling~\citep{ebener-margazoglou-friedrich-etal:2019}. We visualize the transition tube
information obtained from the eigenvalues and eigenfunctions
of the projected second variation operator.
The upper left subfigure shows the histogram of the full data set for all
times. The
remaining subfigures
show histograms of the transition paths at specific
times $t$. The black lines, as a comparison, are the level sets of the
normal PDF with covariance $\eps {\cal C}_z(t,t)$, found by
evaluating~\eqref{eq:cov-eig} numerically, and mean $\phi_z(t)$. Note
that the deformation of the distribution of $X_t^\eps$, conditioned
on $f(X_T^\eps) = z$, is captured quite well using the quadratic,
sampling-free approximation.}
\label{fig:2mb-tube}
\end{figure*}

As in the previous sections, we use the model SDE~\eqref{eq:2d-example}
with $z = 3$ and $\eps = 0.5$ to
illustrate these findings. To do this, we compare the PDF of $X_t^\eps$
at different times $t$, when conditioning on $f(X_T^\eps) = z$, as obtained via
sampling, to the Gaussian approximation ${\cal N}(\phi_z(t), \eps {\cal C}_z(t,t))$
that we evaluate using the instanton as well as eigenvalues and
eigenfunctions of $A_z$
that were computed previously. We use instanton-based importance
sampling~\citep{ebener-margazoglou-friedrich-etal:2019} to
generate $10^5$ trajectories of~\eqref{eq:2d-example} that
satisfy $f(X_T^\eps) = z$ up to a given precision $f((X_T^\eps - \phi_z(T))/\sqrt{\eps}) <
0.05$; the corresponding code, which again uses Euler steps with an integrating factor and a step
size of $\Delta t = 5 \cdot 10^{-4}$,
can be found in the GitHub repository
\citep{Schorlepp-github}. 
Essentially,
instead of using~\eqref{eq:sde} directly, we shift the system by the instanton
(cf.~\cite{tong-vanden-eijnden-stadler:2021} for a visualization 
and further analysis), solve
\begin{align}
\dd Y_t^\eps = \frac{b\left(\phi_z(t) + \sqrt{\eps} Y_t^\eps\right)
- b(\phi_z(t))}{\sqrt{\eps}} \dd t + \sigma \dd B_t\,, \quad Y_0^\eps = 0\,,
\end{align}
and reweight the samples by
\begin{align}
&\exp \bigg\{ \eps^{-1} \int_0^T \big \langle b\left(\phi_z(t) +
\sqrt{\eps} Y_t^\eps\right) - b(\phi_z(t)) \nonumber\\
&- \sqrt{\eps} \nabla
b(\phi_z(t)) Y_t^\eps ,\theta_z \big \rangle_n \dd t 
+ \eps^{-1} \lambda \big(f\left(\phi_z(T) + \sqrt{\eps} Y_T^\eps\right) \nonumber\\
&\qquad- f(\phi_z(T)) - \sqrt{\eps} \nabla f(\phi_z(T)) Y_T^\eps \big) \bigg\}.
\end{align}
The results are shown in figure~\ref{fig:2mb-tube}, and we observe
good agreement between the sampled conditioned distributions
at times $t \in \{0.05, 0.25, 0.5, 0.75, 0.95 \}$ and the corresponding theoretical
small-noise Gaussian approximations. In particular, the deformation of
the fluctuation PDF along the instanton trajectory $\left(\phi_z(t)\right)_{
t \in [0,T]}$ is captured by the Gaussian approximation. It is not surprising that
the Gaussian approximation works well for the parameters $\eps, z$ and
$T$ used here, since the probability $P_F^\eps(z)$ in
section~\ref{subsec:laplace-finite} as approximated by the Laplace method
also matched the direct sampling estimate.

\section{Computational examples}
\label{sec:ex}

We now apply the numerical methods introduced in the previous section to two
high-dimensional examples involving SPDEs: In section~\ref{subsec:kdv},
we consider the Korteweg--De Vries equation in one spatial dimension,
subject to spatially smooth Gaussian noise, and compute precise estimates
for the probability to observe large wave heights at one instance in space
and time. We compare our asymptotically sharp estimates to direct sampling,
and also explicitly compare the two different prefactor computation strategies.
Then, we focus on the stochastically forced three-dimensional incompressible
Navier--Stokes equations in section~\ref{subsec:nse}.
This is a much higher-dimensional problem, and we demonstrate that the
eigenvalue-based prefactor computation indeed remains applicable in practice for
this example. Note that both SPDE examples in this section have periodic
boundary conditions in space, but this is not a restriction of the method
and has merely been chosen for convenience.

\subsection{Stochastic Korteweg--De Vries equation}
\label{subsec:kdv}

\begin{figure*}
\centering
\includegraphics[width = \textwidth]{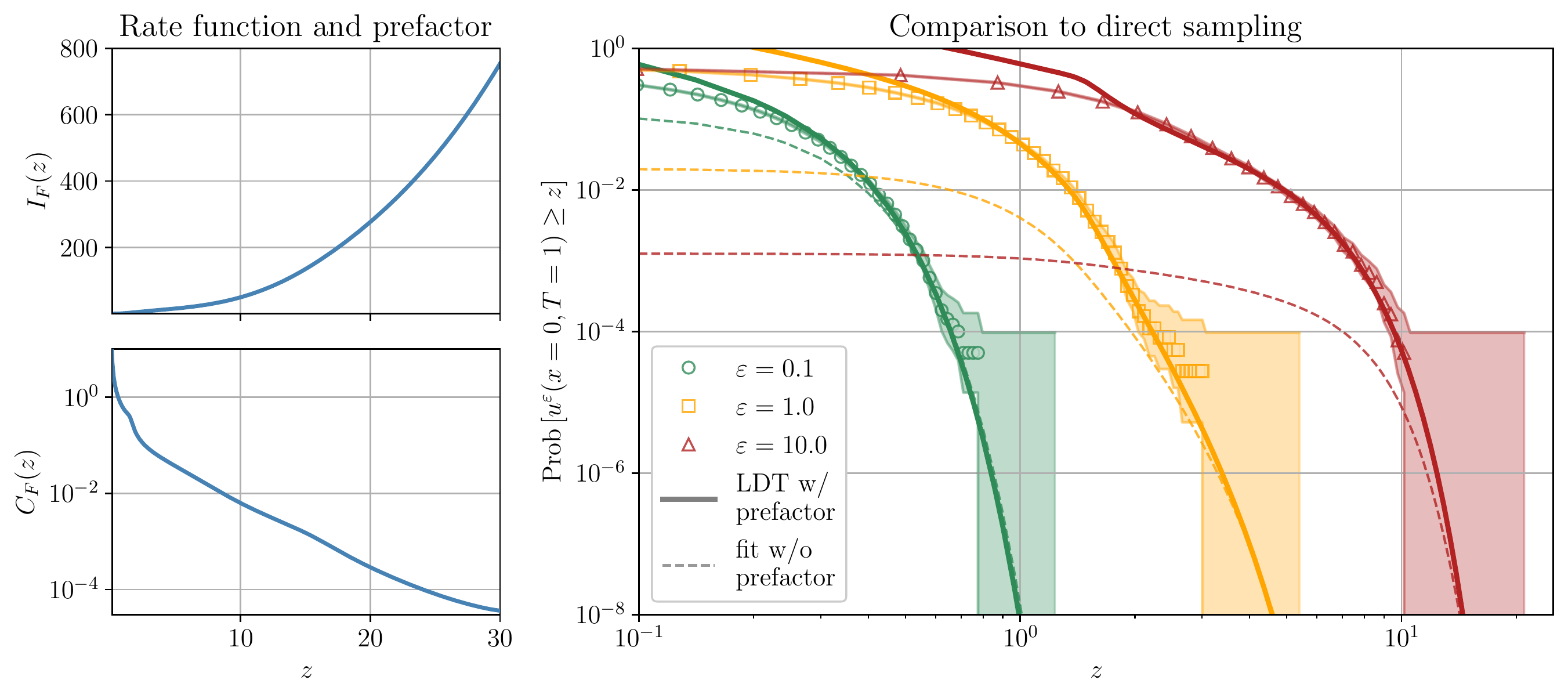}
\caption{Left column: Rate function $I_F$ (top)
and leading order prefactor $C_F$ (bottom) for
the KdV equation~\eqref{eq:kdv} with height observable~\eqref{eq:kdv-obs}, as
obtained from numerical instanton and prefactor computations.
Note that the prefactor depends strongly, almost exponentially,
on the observable value $z$ in this example.
Right: Comparison
of LDT estimate~\eqref{eq:extreme-event-finite-dimensional-result}
for different noise levels $\eps \in \{0.1, 1, 10\}$ to
direct sampling for the SPDE~\eqref{eq:kdv}. For each $\varepsilon$, we
computed $4 \cdot 10^4$ samples of $f\left(u^\eps(\cdot, T) \right)$
to estimate the tail probabilities for various $z$.
The shaded regions are $95\%$ Wilson score intervals~\citep{brown-cai-dasgupta:2001} for
the sampling estimate of the tail probabilities. The solid lines show
the asymptotically sharp
estimate~\eqref{eq:extreme-event-finite-dimensional-result}
without adjustable parameters. In comparison to this,
the dashed lines show just the leading order LDT term $\exp\left\{-I_F(z) /
\eps \right\}$ with a \textit{constant} prefactor (chosen such that the curve matches~\eqref{eq:extreme-event-finite-dimensional-result} for large $z$), which shows that
the prefactor $C_F$ is absolutely necessary to get useful results in this
example at $\eps > 0.1$ and can be understood in regard to the left
column of the figure. Results use $n_x = 1024, n_t = 4000$ for
the instanton computations, pseudo-spectral code with integrating factor,
L-BFGS optimization with penalty term for observable; 80 eigenvalues with
largest absolute value for Fredholm determinant; stochastic Heun steps
with size $\Delta t = 10^{-3}$ for direct sampling.}
\label{fig:kdv-prefac-prob}
\end{figure*}

To illustrate the instanton and prefactor computation,
we study the Korteweg--De Vries (KdV) equation subject to large-scale
smooth Gaussian noise. The KdV equation can be considered as a
model for shallow water waves, so the problem we are interested
in is to estimate the probability of observing large wave amplitudes.
Since this is the first PDE example we study and the general theory
in the previous sections has only been developed for ODEs, we explicitly state the
instanton equations, second order adjoint equations and Riccati equation.
We consider a field $u^\eps \colon [0,l = 2 \pi] \times [0,T = 1] \to \RR$
with periodic boundary conditions in space satisfying the SPDE
\begin{align}
\begin{cases}
\partial_t u^\eps + u^\eps \partial_x u^\epsilon
- \nu \partial_{xx} u^\eps + \kappa \partial_{xxx} u^\eps = \sqrt{\eps}\eta\,,\\
u^\eps(\cdot, 0) = 0\,,
\end{cases}
\label{eq:kdv}
\end{align}
with constants $\nu = \kappa = 4 \cdot 10^{-2}$ and white-in-time,
centered and stationary Gaussian forcing
\begin{align}
\EE \left[\eta(x,t)\eta(x',t') \right] = \chi(x-x') \delta(t-t') \,.
\end{align}
We choose $\hat{\chi}_k = \delta_{\abs{k},1} / (2 \pi)$ as the spatial
correlation function of the noise $\eta$ in Fourier space, with
$\,\hat{}\,$ denoting the spatial Fourier transform. Concretely,
$\eta(x,t)$ is then given by $\eta(x,t) = \pi^{-1/2} (\dot{B}_1(t)
\sin(x) + \dot{B}_2(t) \cos(x))$, where $B_1,B_2$ are independent
standard one-dimensional Brownian motions.
Hence, the forcing only acts on a single large scale
Fourier mode, and excitations of all other modes are due to the
nonlinearity of the SPDE. As our observable, we choose the wave height
at the origin
\begin{align}
f(u(\cdot, T)) = u(0,T)\,,
\label{eq:kdv-obs}
\end{align}
and we want to quantify the tail probability $P_F^\eps(z)
= \PP \left[f(u^\eps(\cdot, T)) \geq z \right]$ for
different $z > 0$.
Note that the effective dimension of the system when formulated in
terms of the noise for our choice of noise correlation is small,
and we have $\rank \sigma = 2 \ll n = n_x$ for typical spatial resolutions.
Unless otherwise specified, we use $n_x = 1024$ for all numerical results in this section, as well as $n_t = 4000$ equidistant points in time,
and we expect the prefactor computation in terms of eigenvalues of
$A_z$ to be more efficient in this example, even though the Riccati
approach still remains feasible.

We use a pseudo-spectral code and explicit second order
Runge-Kutta steps in time with an integrating factor for the linear terms.
The final-time constraint is treated with the augmented Lagrangian method.
Denoting the state space instanton by $u_z$ with adjoint variable
$p_z$ and Lagrange multiplier $\lambda_z$, the first-order necessary
conditions at the minimizers read
\begin{align}
  &\begin{cases}
    \partial_t u_z =-u_z \partial_x u_z + \nu \partial_{xx} u_z
    - \kappa \partial_{xxx} u_z + \chi * p_z\,,\\
    \partial_t p_z = -u_z \partial_x p_z - \nu \partial_{xx} p_z
    - \kappa \partial_{xxx} p_z \,,
  \end{cases}\nonumber\\
  \text{with }&\begin{cases}
  u_z(\cdot, 0) = 0\,, \quad f(u_z(\cdot, 1)) = z\,,\\
  p_z(x, 1) = \lambda_z \delta(x)\,.
  \end{cases}
\end{align}
Here, $*$ denotes spatial convolution, which appears
due to the stationarity of the forcing.

As a starting point, we compute instantons for a range of equidistantly
spaced observable values $z \in [0, 30]$.
Knowledge of the
instanton for different $z$ gives us access to the rate
function $I_F$ of the observable, which is shown on the left in
figure~\ref{fig:kdv-prefac-prob}.

\begin{figure}
\centering
\begin{ruledtabular}
\begin{tabular}{cccc}
resolution $(n_x, n_t)$ & \#iterations & $I_F(z)$ \\
\hline
$(32, 125)$ & $286$ & $44.106$\\
$(64, 250)$ & $310$ & $34.787$\\
$(128, 500)$ & $268$ & $34.605$\\
$(256, 1000)$ & $283$ & $34.681$\\
$(512, 2000)$ & $255$ & $34.694$\\
$(1024, 4000)$ & $259$ & $34.696$
\end{tabular}
\end{ruledtabular}\\
\vspace{.3cm}
\includegraphics[width = .48 \textwidth]{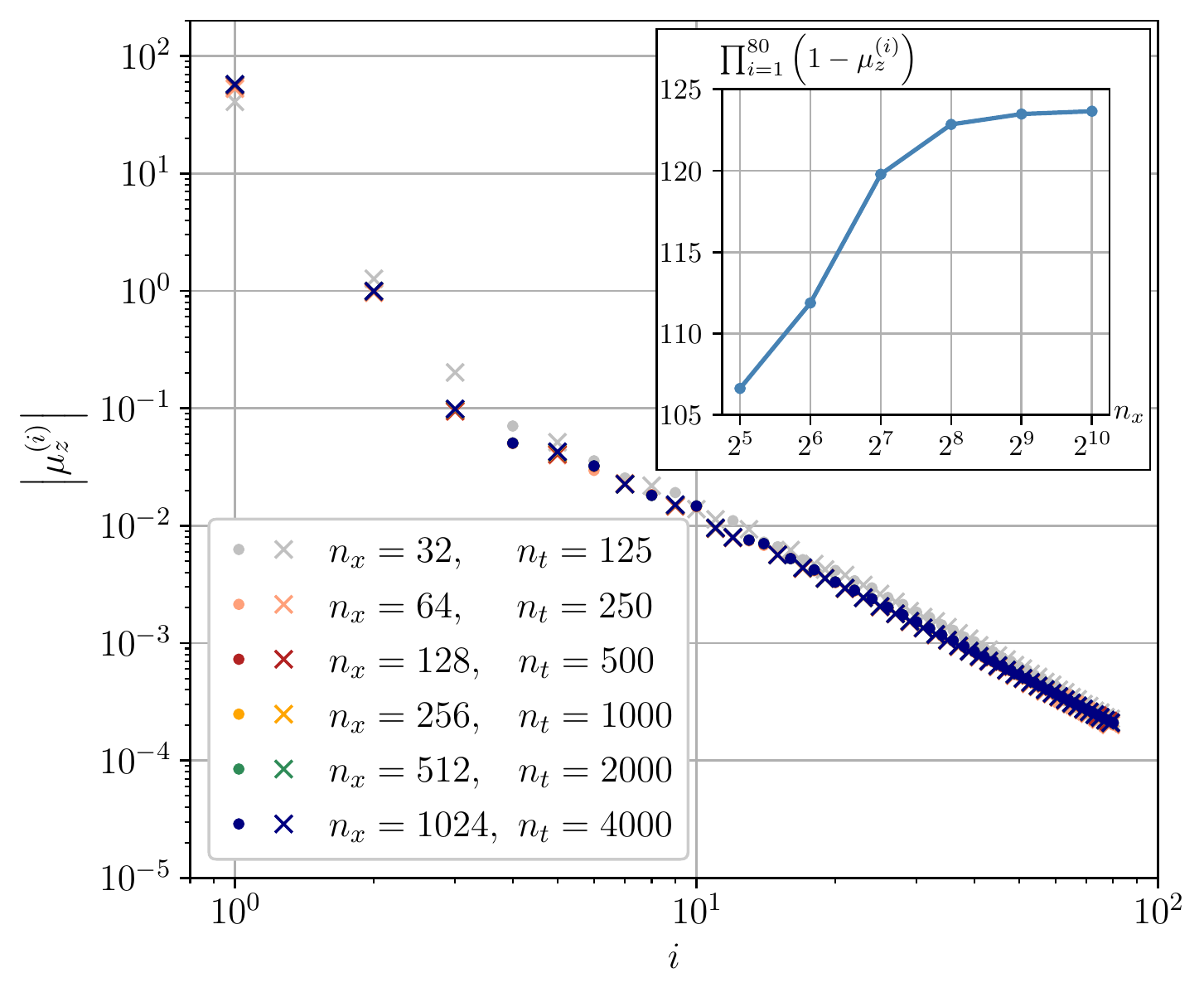}
\caption{Performance of instanton and prefactor computations for KdV
  problem with $z=8.4$ for different spatio-temporal resolutions
  $(n_x, n_t) \in \{(32, 125), \dots, (1024, 4000)\}$.  The table
  shows the number of optimization iterations required to compute the
  instanton, and the value of the objective $I_F(z)$. The number of
  iterations does not increase with the resolution $(n_x,
  n_t)$.
  The bottom figure shows 80 eigenvalues
  $\mu_z^{(i)}$ with largest absolute value of $A_z$. The main figure
  shows the absolute value of the eigenvalues $\mu_z^{(i)}$ (dots:
  positive eigenvalues, crosses: negative eigenvalues). The inset
  shows $\prod_{i = 1}^{80} \left(1 - \mu_z^{(i)} \right)$ for the
  different resolutions $(n_x, n_t)$ as an approximation for the
  Fredholm determinant $\det (\Id - A_z)$, which is seen to converge
  with increasing resolution.  Note that only for the lowest
  resolution, the eigenvalue spectrum shows noticeable deviations from
  the results at $(n_x, n_t) = (1024, 4000)$. The latter resolution
  has been used for all other numerical results on the KdV equation in
  this paper.}
\label{fig:kdv-evals-diff-res}
\end{figure}

In the table in figure~\ref{fig:kdv-evals-diff-res}, we
show for fixed $z$ how the value of $I_F(z)$ converges
when increasing the spatio-temporal resolution, and in particular that
the number of optimization steps needed to find the instanton is robust
under changes of the numerical resolution, indicating scalability of
the instanton computation. The numerical details for
these instanton computations are as follows
(cf.~\cite{schorlepp-grafke-may-etal:2022}): Initial control $p
\equiv 0$ and initial Lagrange multiplier $\lambda = 0$; precise
target observable value $z = 8.39125$; $6$ logarithmically spaced penalty
steps from $1$ to $300$ for augmented Lagrangian method;
optimization is terminated upon reduction of
gradient norm by $10^6$; same (presumably) global
minimizer was found for each resolution;
discretize-then-optimize; L-BFGS solver with $4$ updates stored;
Armijo line search.

\begin{figure*}
\centering
\includegraphics[width = \textwidth]{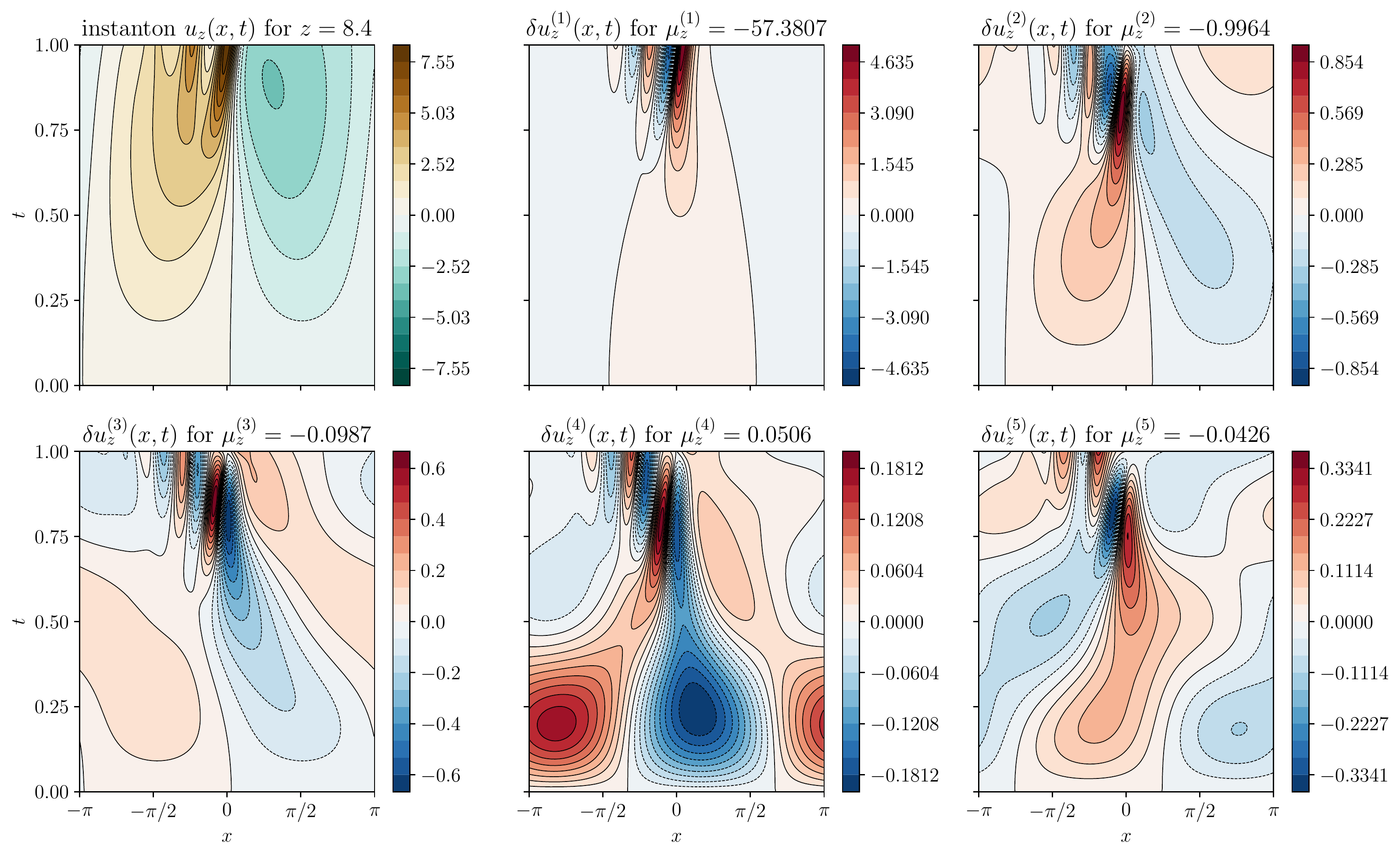}
\caption{Example instanton field~$u_z$ in space and time for~$z = 8.4$
(top left) for the KdV equation~\eqref{eq:kdv} and height
observable~\eqref{eq:kdv-obs}, and dominant~$5$ normalized state variable
eigenfunctions~$\delta u_z^{(i)}$ of the projected second variation
operator~$A_z$. Due to the KdV nonlinearity and linear wave dispersion,
the large-scale forcing input is transformed into a large wave with dominant
peak at $t = T$, $x = 0$ for the instanton $u_z$, i.e.\ the most likely
field realization to obtain a large wave height $z = 8.4$ at $t = T$, $x = 0$.
The strongest fluctuations around the instanton resemble the instanton itself, but
are necessarily centered around $0$ with final-time
height $\delta u(0, T) = 0$ at the origin. Note that only two eigenvalues
are larger than $0.1$ in modulus, reflecting the small effective dimension
of the system in the noise variable, and that $\delta u_z^{(4)}$ contains
already higher modes in time.}
\label{fig:kdv-gammas}
\end{figure*}

Two comments on the instanton computations
for this example are in order: Firstly, the observable rate function is
non-convex for some~$z$ in the interval $[1.5, 5]$ (not visible in the figure). This poses a problem
for the dual problem solved at fixed $\lambda$ without penalty, but is not an
issue for the penalty or augmented Lagrangian strategy that we used.
Furthermore, this means that the Riccati prefactor computation is not
directly applicable in this region, but the Fredholm expression remains valid.
Secondly, since it is a priori unclear whether the minimization problem
for the instanton has a unique solution (the target functional is quadratic,
but the constraint is nonlinear), we started multiple optimization runs for the
same~$z$ at different random initial conditions. In the KdV system, we found
multiple subdominant minima that consist of multiple large wave
crests (as opposed to just one for the dominant one, as shown in the
top left of figure~\ref{fig:kdv-gammas} for one~$z$), but only
took the (presumably) global minimizer for subsequent estimates.

\begin{figure}
\centering
\includegraphics[width = .48 \textwidth]{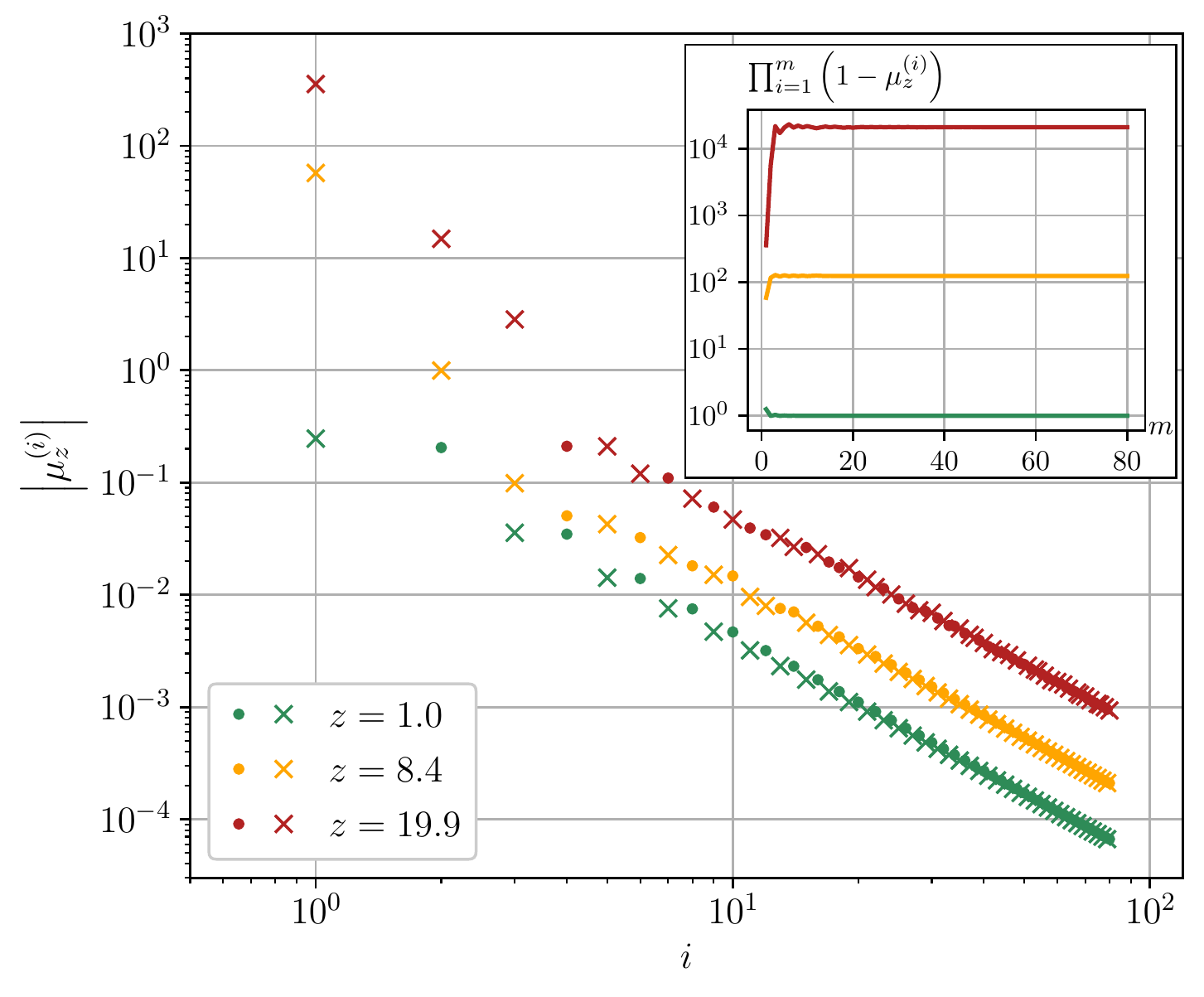}
\caption{Result of numerically computing 80 eigenvalues $\mu_z^{(i)}$ with largest
absolute value of $A_z$ for the KdV equation~\eqref{eq:kdv}
with $z \in \{1, 8.4, 19.9\}$. Main figure: absolute value of
the eigenvalues $\mu_z^{(i)}$ (dots: positive eigenvalues,
crosses: negative eigenvalues). Inset: Finite product $\prod_{i = 1}^m
\left(1 - \mu_z^{(i)} \right)$ for different $m$ as an approximation for
the Fredholm determinant $\det (\Id - A_z)$. We see that the
eigenvalues rapidly decay to zero  for all $z$. Similarly,
the cumulative product in the inset converges quickly, and
the determinant is in fact well-approximated by less than 10 eigenvalues
 for all $z$.}
\label{fig:kdv-evals}
\end{figure}

To complete the asymptotic estimate of the wave height probability
via~\eqref{eq:extreme-event-finite-dimensional-result}, we further need
the prefactor $C_F(z)$ for all $z$, which we compute by finding the
dominant
eigenvalues of $A_z$ as before.
We specify the input and output
of the linear operator $A_z$ only in terms of the two real Fourier modes of the
noise that are relevant for this, to remove the memory cost
of the eigenvalue solver. The second order adjoint equations~\eqref{eq:second-order-adj-eq} for noise fluctuations $\delta \eta \colon [0,2 \pi] \times [0,1] \to \RR$ for the KdV equation read
\begin{align}
  &\begin{cases}
    \partial_t \delta u =-\partial_x (u_z \delta u) + \nu \partial_{xx}
    \delta u - \kappa \partial_{xxx} \delta u + \chi^{1/2} * \delta \eta\,,\\
    \partial_t \delta p = -\delta u \partial_x p_z -u_z \partial_x \delta p
    - \nu \partial_{xx} \delta p - \kappa \partial_{xxx} \delta p\,,
  \end{cases} \nonumber\\
  \text{with }&\begin{cases}
  \delta u (\cdot, 0) = 0\,,\\
  \delta p(\cdot, 1) = 0\,,
  \end{cases}
  \label{eq:scd-order-kdv}
\end{align}
with $A_z \delta \eta =  \chi^{1/2} * \delta p$. In our implementation, we supply
the second variation operator with the two real Fourier
coefficients $\left(\text{Re} \, \widehat{\delta \eta}_1(t_i)\right)_{i
= 0, \dots, n_t}$ and $\left(\text{Im} \, \widehat{\delta \eta}_1(t_i)
\right)_{i = 0, \dots, n_t}$, assemble the full fluctuation
vector $\delta \eta$ from it, and return $\chi^{1/2} * \delta p$ in the
same format after solving~\eqref{eq:scd-order-kdv}. As the KdV solutions
fit into memory, checkpointing, as discussed
in section~\ref{subsec:pref-fred-adj-comp}, is not necessary. In
figure~\ref{fig:kdv-evals}, we show the convergence
of the determinant $\det(\Id - A_z)$ for some $z$'s based on the
found eigenvalues, thereby demonstrating that  a handful of eigenvalues
suffices for an accurate approximation of the prefactor. The number of
necessary eigenvalues increases only weakly with the observable
value $z$ in this example. In addition,
figure~\ref{fig:kdv-evals-diff-res} shows the effect of
varying the spatio-temporal resolution $(n_x, n_t)$ on the
determinant $\det(\Id - A_z)$ for one particular observable value
of $z = 8.4$ at a fixed number of computed eigenvalues. We see that
as long as the physical problem is resolved, the eigenvalue spectrum does
not change much with the resolution, and the determinant converges when
increasing the spatio-temporal resolution. This indicates that our
methods are scalable, i.e., their cost does not increase with the
temporal (and also spatial) discretization beyond the
increased  cost of the PDE solution. This is a crucial property of the
eigenvalue-based prefactor computation and is in contrast with the
Riccati approach.

The result for the prefactor $C_F$ as a function of $z$ is shown
on the bottom left of figure~\ref{fig:kdv-prefac-prob}. Note that
the vertical axis is scaled logarithmically, i.e.\ the prefactor strongly
depends on the observable value. The importance of the prefactor
is further confirmed by the comparison of the complete asymptotic
estimate~\eqref{eq:extreme-event-finite-dimensional-result} to the results
of direct Monte Carlo simulations on the right in
figure~\ref{fig:kdv-prefac-prob}. For
three values of $\eps \in \{0.1, 1, 10\}$, we performed $4 \cdot 10^4$
respective simulations of the stochastic KdV equation~\eqref{eq:kdv}
to estimate the tail probability $P_F^\eps(z)$ without
approximations. 
Using both the rate function and prefactor, excellent agreement with the
Monte Carlo simulations is obtained. In contrast to this, only using the
leading order LDT term $\exp \left\{-I_F(z)/\eps \right\}$ with a constant
prefactor leads to a much worse agreement with simulations, and in fact
only works reasonably for $\eps = 0.1$. Note also that one can see
from these comparisons that the actual effective smallness parameter for
the asymptotic expression~\eqref{eq:extreme-event-finite-dimensional-result}
to be valid is $\eps / h(z)$ for some monotonically increasing function $h$,
meaning that the estimate is also valid for large $\eps$ as long as suitably
large $z \to \infty$ are considered. In this sense, the estimate is truly an
extreme event probability estimate, but we chose to work in terms
of the formal parameter $\eps$ to have an explicit and general scaling parameter,
in contrast to the example-specific function $h(z)$. For works on
large deviation principles directly in $z \to \infty$, see
e.g.~\cite{dematteis-grafke-vanden-eijnden:2019,tong-vanden-eijnden-stadler:2021}

In addition to the probability estimate itself, the instanton,
eigenvalues and eigenfunctions of $\eta_z$ also carry physical information
about the system, as discussed in general in section~\ref{sec:prob-interp}.
Figure~\ref{fig:kdv-gammas} shows the instanton $u_z$, i.e.\ the most likely
field realization to reach a large wave height of $z = 8.4$, and the
dominant space-time fluctuations $\delta u_z^{(i)}$ around it.

\begin{figure*}
\centering
\includegraphics[width = \textwidth]{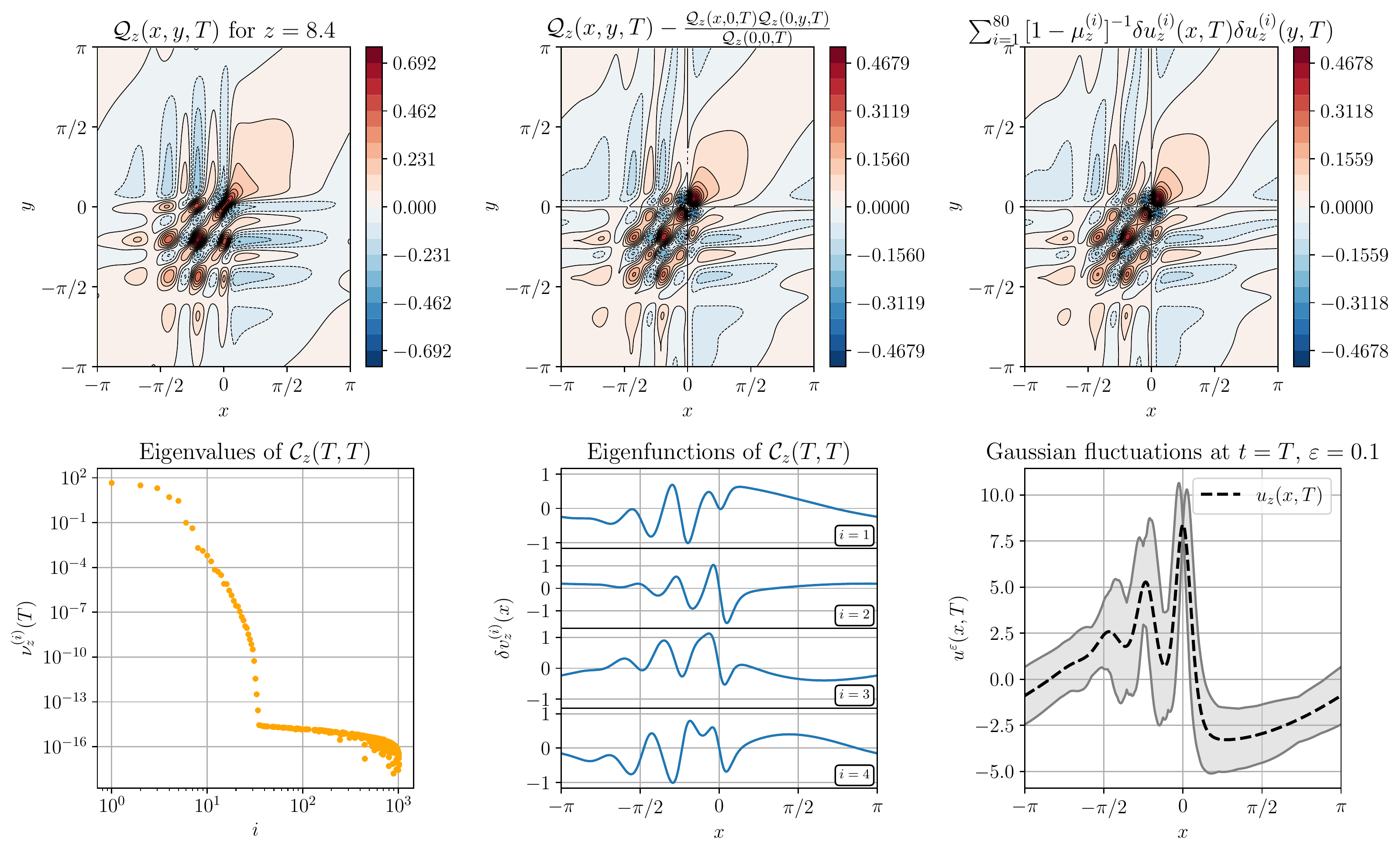}
\caption{Information on the conditioned final time Gaussian fluctuations
around the KdV instanton for $z = 8.4$, calculated from the quantities used
to evaluate the prefactor $C_F(z)$. Top, left: Riccati solution ${\cal Q}_z(
\cdot, \cdot, T = 1)$ at final time. Top, center: Projection of the Riccati
solution, such that the constraint $\delta u(0, T) = 0$ is satisfied. This way,
the final time covariance ${\cal C}_z(T,T)$ as given in~\eqref{eq:cov-ric} is
obtained. Top, right: The same final time covariance ${\cal C}_z(T,T)$
constructed from the eigenvalues and eigenfunctions of $A_z$ instead as
in~\eqref{eq:cov-eig}. The result is visually indistinguishable from the Riccati
computations. Bottom, left: Eigenvalues $\nu_z^{(i)}(T)$ of the
covariance ${\cal C}_z(T,T)$. We see that the eigenvalues quickly decay to
zero, and less than 10 fluctuation modes are in fact relevant. Bottom,
center: Eigenfunctions $\delta v_z^{(i)}$ for the $4$ dominant
eigenvalues $\nu_z^{(i)}(T)$, $i \in \{1,2,3,4\}$, which all necessarily
satisfy $\delta v_z^{(i)}(x = 0) = 0$. Bottom, right:
Instanton $u_z(\cdot, T)$ at final time (dashed line), and variance of
conditioned Gaussian fluctuations around it for $\eps = 0.1$ (shaded area).}
\label{fig:kdv-final-time}
\end{figure*}

We further computed the Gaussian fluctuations around the instanton for $z = 8.4$ at
the final instance $t = T$ in figure~\ref{fig:kdv-final-time}.
First of all, we also solved the forward Riccati equation~\eqref{eq:riccati-fw},
which is a PDE for ${\cal Q}_z \colon [0,2 \pi]^2 \times [0,1] \to \RR$ here
and reads
\begin{align}
\begin{cases}
 \partial_t {\cal Q}_z(x, y, t) = \chi(x - y)\\
 \hspace{2cm}- \left[\partial_x \left(u_z(x) \cdot \right)
 + \partial_y \left(u_z(y) \cdot \right) \right]{\cal Q}_z(x, y, t)\\
 \hspace{2cm}+ \nu \left[\partial_{xx} + \partial_{yy} \right]{\cal Q}_z(x, y, t) \\
 \hspace{2cm}- \kappa \left[\partial_{xxx}
 + \partial_{yyy} \right]{\cal Q}_z(x, y, t)\\
 \hspace{2cm}+ \int_0^{2 \pi} {\cal Q}_z(x, x', t) 
 \partial_{x'} p_z(x', t){\cal Q}_z(x', y, t) \dd x'\,,\\
 {\cal Q}_z(\cdot, \cdot, t = 0) = 0\,,
\end{cases}
\end{align}
using the same pseudospectral code and explicit second order Runge-Kutta steps
with integrating factor. The result for the prefactor agrees with the one
obtained using the Fredholm determinant expression, with $C_F(z = 8.4)
\approx 1.0793 \cdot 10^{-2}$ using the eigenvalues and $C_F(z = 8.4)
\approx 1.0794 \cdot 10^{-2}$ from the Riccati approach with
\begin{align}
C_F(z) = \frac{\exp \left\{\tfrac{1}{2} \int_0^1 \dd t \int_0^{2 \pi}
\dd x \; \partial_x p_z(x, t) {\cal Q}_z(x,x,t) \right\}}{\lambda_z
\sqrt{{\cal Q}_z(0,0,1)}}\,.
\end{align}
For this particular observable value, the Riccati equation could be integrated
without numerical problems, but we encountered a removable singularity for larger
observable values. The final-time covariance of the conditioned
Gaussian fluctuations around the instanton, as predicted using either the
Riccati solution~\eqref{eq:cov-ric} or the eigenfunctions and
eigenvalues~\eqref{eq:cov-eig}, indeed coincides for both approaches and
is highly oscillatory (top row, center and right in
figure~\ref{fig:kdv-final-time}). Denoting the eigenvalues and
normalized eigenfunctions of the final-time covariance
operator ${\cal C}_z(T,T)$ by $\nu_z^{(i)}(T)$ and $\delta v_z^{(i)}$,
we see that only a handful of fluctuation modes~$\delta v_z^{(i)}$ are
actually observable since the
eigenvalues~$\nu_z^{(i)}(T)$ in the bottom left of
figure~\ref{fig:kdv-final-time} quickly decay. Using the eigenvalues
and eigenfunctions, realizations
of $u^\eps(\cdot, T)$ when conditioning on $u^\eps(0, T) = z = 8.4$ can
now easily be sampled within the Gaussian approximation as
\begin{align}
u^\eps(x, T) \approx u_z(x, T) + \sqrt{\eps}
\sum_{i = 1}^\infty Z_i  \sqrt{\nu_z^{(i)}(T)} \delta v_z^{(i)}(x)
\end{align}
with $Z_i$ independent and identically standard normally distributed.
All in all, this example demonstrates the practical relevance and ease of
applicability of the asymptotically sharp LDT estimate including the prefactor
in a nonlinear, one-dimensional SPDE.

\subsection{Stochastically forced incompressible three-dimensional Navier--Stokes equations}
\label{subsec:nse}

As a challenging, high-dimensional example, we consider the estimation
of the probability of a high strain event
in the stochastically forced incompressible three-dimensional
Navier--Stokes equations. Our main goal here is to demonstrate that in
addition to instantons for this problem, which were computed
by~\cite{schorlepp-grafke-may-etal:2022}, it is also numerically feasible
to compute the leading order prefactor using the Fredholm determinant
approach~\eqref{eq:tail-prob-prefac-sde}.
Our setup hence follows the one treated
by~\cite{schorlepp-grafke-may-etal:2022}.
A comprehensive analysis of the problem, including the behavior of
the prefactor in the vicinity of the critical points of the
dynamical phase transitions observed
in this example, is beyond the scope of this paper.
For other works on instantons and large deviations for the
three-dimensional stochastic Navier--Stokes equations,
see~\cite{falkovich-kolokolov-lebedev-etal:1996,moriconi:2004,grafke-grauer-schaefer:2015,apolinario-moriconi-pereira-etal:2022}. We consider a
velocity field $u^\eps \colon [0,l = 2 \pi]^3 \times [0,T = 1] \to \RR^3$
with periodic boundary conditions in space that satisfies
\begin{align}
\begin{cases}
\partial_t u^\eps + \left(u^\eps \cdot \nabla \right) u^\epsilon
- \Delta u^\eps + \nabla P = \sqrt{\eps}\eta\,,\\
\nabla \cdot u^\eps = 0\,,\\
u^\eps(\cdot, 0) = 0\,.
\end{cases}
\label{eq:nse}
\end{align}
Here, $P$ denotes the pressure which is determined through
the divergence constraint.
The forcing $\eta$ is centered Gaussian, large-scale in space,
white in time, and solenoidal with covariance
\begin{align}
\EE \left[\eta(x,t)\eta(x',t')^\top \right] = \chi(x-x') \delta(t-t') \,,
\end{align}
where a Mexican hat correlation function with correlation length $1$
\begin{align}
\chi(x) = \left[ 1_{3 \times 3} - \frac{1}{2} \left(
\norm{x}^2 1_{3 \times 3} - x \otimes x\right) \right]
\exp \left\{-\frac{\norm{x}^2}{2} \right\}\,,
\end{align}
is used. Note that this corresponds to the
situation $\rank \sigma \ll 3 n_x^3$
of section~\ref{subsec:pref-fred-adj-comp}, where only a
small number of degrees of freedom is forced due
to the Fourier transform $\hat{\chi}$ decaying exponentially.
As our observable, we consider the strain $f(u) = \partial_3 u_3(x=0)$ at
the origin. Denoting the Leray projection onto the divergence-free part of
a vector field by ${\cal P}$, the instanton equations
for $(u_z, p_z, \lambda_z)$ are given by
\begin{align}
  &\begin{cases}
    \partial_t u_z =- {\cal P} \left[ \left(u_z \cdot \nabla \right)
    u_z \right] + \Delta u_z + \chi * p_z\,,\\
    \partial_t p_z = - {\cal P} \left[ \left(u_z \cdot \nabla \right)
    p_z  + \left(\nabla p_z \right)^\top u_z\right] - \Delta p_z \,,
  \end{cases}\nonumber\\
  \text{with }&\begin{cases}
  u_z(\cdot, 0) = 0\,, \quad f(u_z(\cdot, 1)) = z\,,\\
  p_z(\cdot, 1) = \lambda_z {\cal P} \left[\left. \fdv{f}{u}
    \right|_{u_z(\cdot, 1)} \right]\,.
  \end{cases}
\end{align}
With the instantons computed, we are able to evaluate
the application of the second variation operator $A_z$ to noise fluctuation
vectors $\delta \eta \colon [0,2\pi]^3 \times [0,1] \to \RR^3$ by solving
the second order adjoint equations
\begin{align}
&\begin{cases}
\partial_t \left(\delta u \right) = -{\cal P} \left[(u_z \cdot \nabla)
\delta u  + (\delta u \cdot \nabla) u_z\right] \\
\hspace{1.4cm}+ \Delta \left(\delta u\right)
+ \chi^{1/2}  * \delta \eta\,,\\
\partial_t \left(\delta p\right) = - {\cal P} \big[ \left(\nabla p_z +
\left(\nabla p_z \right)^\top \right) \delta u + \left(u_z \cdot \nabla
\right) \delta p \\
\hspace{1.4cm}+ \left(\nabla (\delta p) \right)^\top u_z \big] -
\Delta \left(\delta p\right)\,,
\end{cases}\nonumber\\
\text{with }&\begin{cases}
\delta u(\cdot, 0) = 0\,,\\
\delta p(\cdot, 1) = 0\,.
\end{cases}
\label{eq:nse-scd-adj}
\end{align}
We focus on $z = -25$ here, where the unique instanton
solution does not break rotational
symmetry~\citep{schorlepp-grafke-may-etal:2022}. Numerically, we use a
pseudo-spectral GPU code with a spatial resolution $n_x = n_y = n_z = 128$,
a temporal resolution of $n_t = 512$,
a nonuniform grid in time with smaller time steps close to $T = 1$, and second order explicit Runge-Kutta
steps with an integrating factor for the diffusion term. 
We truncated $\chi$ in Fourier space by setting it to 0 for all $k$ where
$\abs{\hat{\chi}_k} < 10^{-14}$, leading to $\norm{k} \leq 9$ and an
effective real spatial dimension, independently of $n_x$, of
approximately $\rank \sigma \approx 2 \cdot (2 \cdot 9)^3 = 11664$
for the noise (by taking a cube instead of sphere for the Fourier coefficients
of the noise vectors that are stored, and noting that $\hat{\chi}_k$ projects
onto $k^\perp$).
The evaluation of the second order adjoint equations is then possible
with only a few GB of VRAM for this resolution when exploiting double
checkpointing and low rank storage as described in
section~\ref{subsec:pref-fred-adj-comp}. We computed the 600 largest
eigenvalues of operator $A_z$, again realized as a
scipy.sparse.linalg.LinearOperator, by using scipy.sparse.linalg.eigs
as before. We transfer the data to the GPU to
evaluate the second variation applied to $\delta \eta$ by
solving~\eqref{eq:nse-scd-adj} with PyCUDA~\citep{kloeckner-etal:2012},
and transfer back
$\chi^{1/2} * \delta p$ to the CPU afterwards. Computing $600$
eigenvalues this way needs about $1200$ operator evaluations, or about
$30$ hours on a modern workstation with Intel Xeon Gold 6342 CPUs at
$2.80\;\text{GHz}$ and an NVIDIA A100 80GB GPU. The main
limitation for computing more eigenvalues is that the eigenvalue
solver used stores all matrix vector products in
RAM. This could be overcome by storing some of them on a
hard disk, or using different algorithms that can be parallelized over
multiple nodes such as randomized SVD~\citep{maulik-mengaldo:2021}.

\begin{figure}
\centering
\includegraphics[width = .48\textwidth]{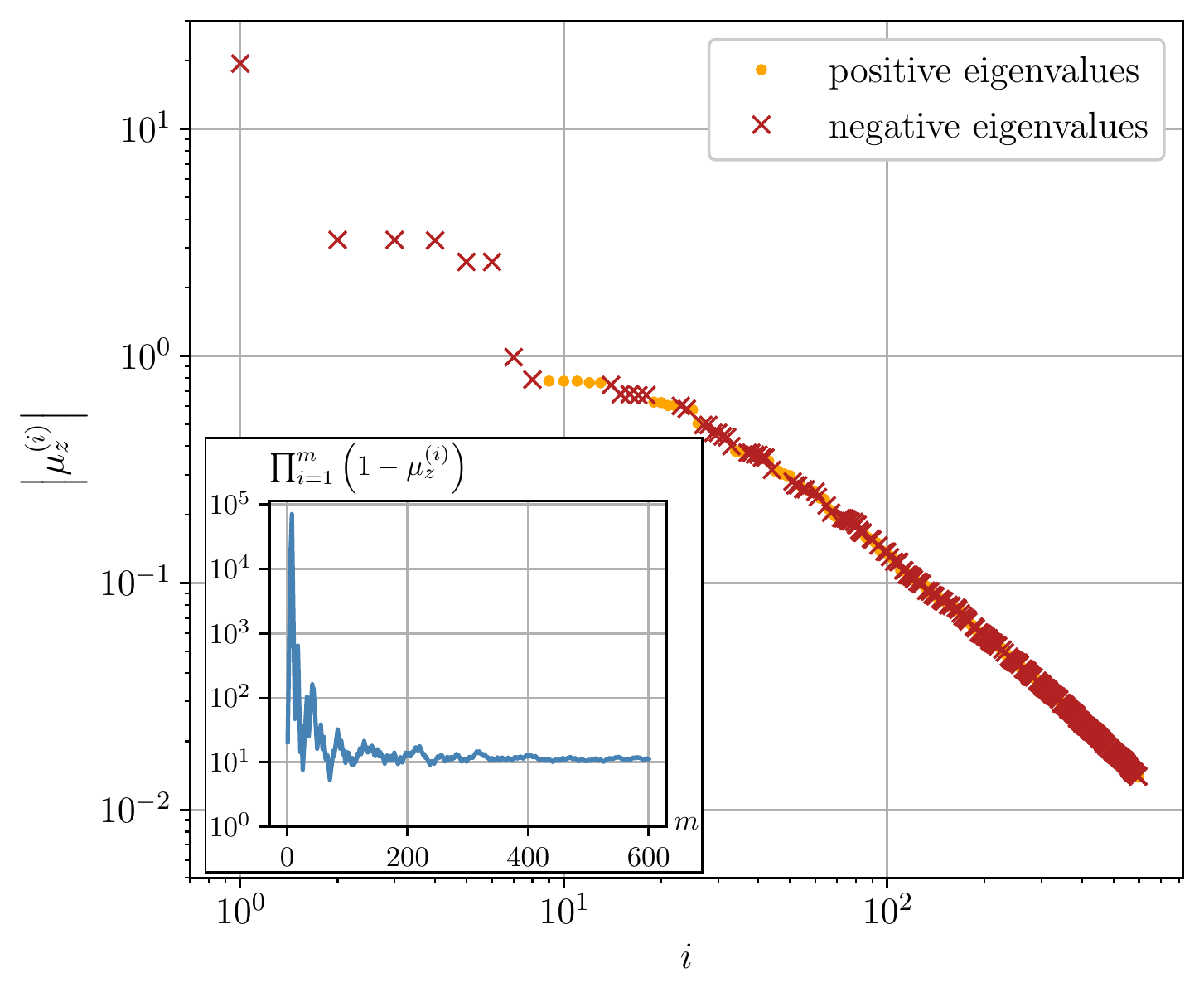}
\caption{Result of numerically computing 600 eigenvalues $\mu_z^{(i)}$ with largest
absolute value of $A_z$ for the three-dimensional Navier--Stokes
equations~\eqref{eq:nse} with strain $z = \partial_3 u_3(x = 0,T) = -25$,
where the instanton is a rotationally symmetric pair of vortex rings.
Main figure: absolute value of
the eigenvalues $\mu_z^{(i)}$. Inset: Finite product $\prod_{i = 1}^m
\left(1 - \mu_z^{(i)} \right)$ for different $m$ as an approximation for
the Fredholm determinant $\det (\Id - A_z)$. We see in the main figure that the eigenvalues often appear in pairs, which
happens whenever the eigenfunctions break the rotational symmetry of the
problem, such that, due to the periodic box, there are two linearly
independent eigenfunctions for the same eigenvalue. The inset shows
that $\det (\Id - A_z)$ is approximately $11$ in this example, but
that even more eigenvalues would be needed to get an accurate result.}
\label{fig:nse-evals}
\end{figure}

\begin{figure*}
\centering
\includegraphics[width = \textwidth]{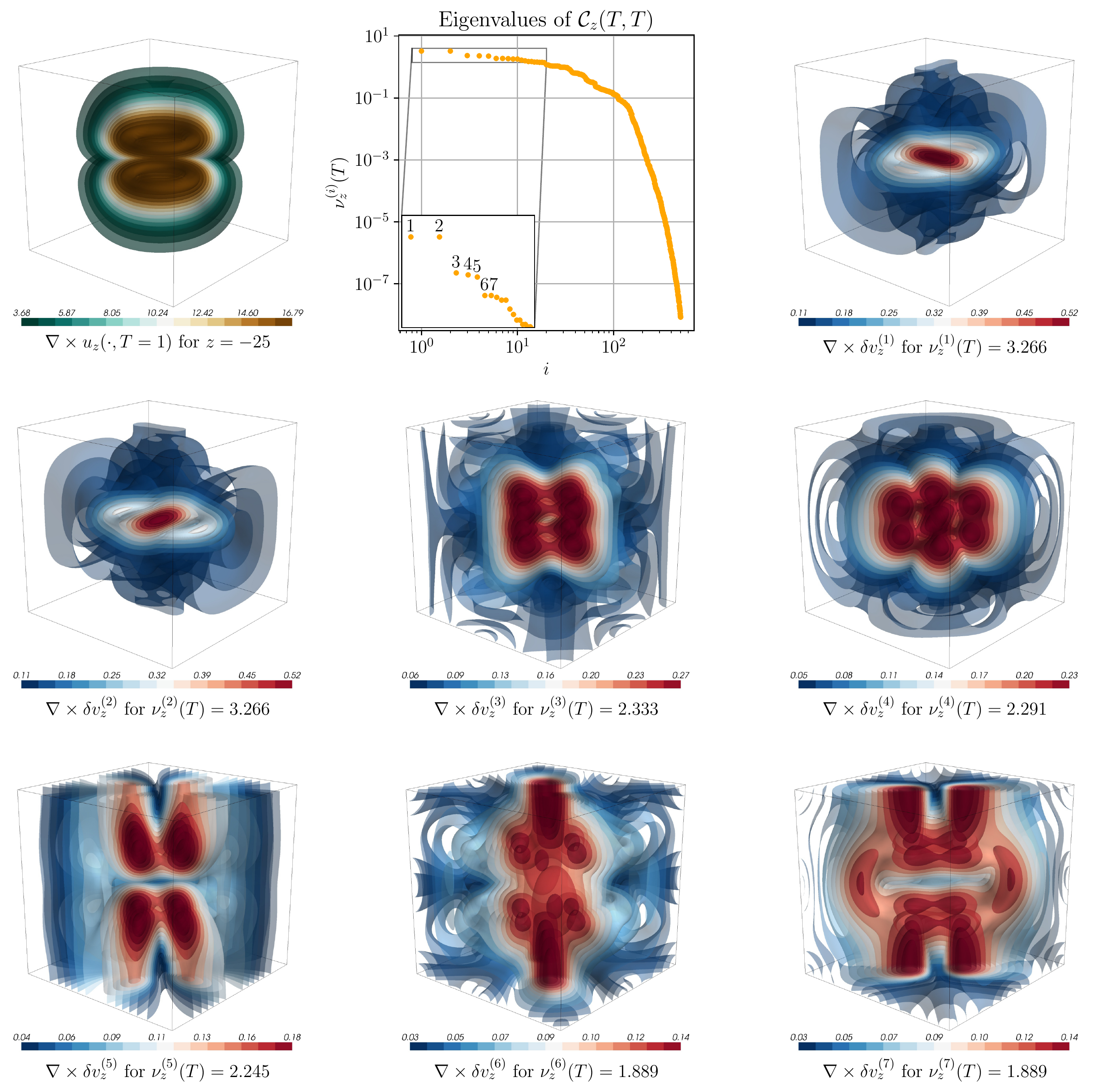}
\caption{Visualization of the instanton and dominant
Gaussian fluctuations around
it at final time $T = 1$, for a strain event
with $z = \partial_3 u_3(x = 0,T) = -25$ for the three-dimensional
Navier--Stokes equations~\eqref{eq:nse}. All three-dimensional images show
isosurfaces of the vorticity or curl of the respective field.
Top, left: The unique instanton
for this observable value is a rotationally symmetric pair of vortex rings.
Top, center: Eigenvalues $\nu_z^{(i)}(T)$ of the final-time covariance
operator ${\cal C}_z(T,T)$, approximated as ${\cal C}_z(T,T)
\approx \sum_{i = 1}^{600} [1 - \mu_z^{(i)}]^{-1}
\delta u_z^{(i)}(\cdot, T) \otimes \delta u_z^{(i)}(\cdot, T)$
using $600$ eigenvalues $\mu_z^{(i)}$ with largest
absolute value of the projected second variation operator $A_z$, of
the conditioned Gaussian fluctuations around the instanton.
Top right, and second/third row: Normalized eigenfunctions $\delta v_z^{(i)}$
of ${\cal C}_z(T,T)$ for the largest eigenvalues of ${\cal C}_z(T,T)$,
indicating the strongest fluctuation directions around the strain instanton
at final time $t = T$.}
\label{fig:nse-evecs}
\end{figure*}

The results for the eigenvalues of $A_z$ are shown in
figure~\ref{fig:nse-evals}. We see that the absolute value of the eigenvalues
decays such that the product $\prod_{i = 1}^m \left(1 - \mu_z^{(i)} \right)$
converges as $m$ increases, but that even more than 600 eigenvalues would be
needed for a more accurate result. For smaller observable values $z$,
faster convergence is expected. Also, the spectrum of $A_z$ shows a
large number of doubly degenerate eigenvalues, which appear whenever
the eigenfunctions break the axial symmetry of the instanton. This feature
of the spectrum clearly depends on the domain and spatial boundary conditions
that were chosen here.
From the instanton computation,
we obtain $I_F(z) \approx 1900.7$ for the rate function, and from
the $600$ eigenvalues of $A_z$ we
estimate $C_F(z) \approx 4.9 \cdot 10^{-3}$. With this, we can estimate that
e.g.\ for $\eps = 250$, the probability to observe a strain event with
$\partial_3 u_3(x = 0,T = 1) \leq -25$ is approximately $1.5 \cdot 10^{-5}$, which
matches the sampling estimate of $P_F^{250}(-25) \in [1.3
\cdot 10^{-5}, 1.7 \cdot 10^{-5}]$ at $95\%$ asymptotic confidence, as obtained
from~$10^4$ direct numerical simulations of~\eqref{eq:nse}
(data set from~\cite{schorlepp-grafke-may-etal:2022}).
For smaller $\eps$, the event becomes more rare, and it quickly becomes
unfeasible to estimate its probability via direct sampling, whereas the
quadratic estimate using the rate function and prefactor can be computed
for any $\eps$ and is known to become more precise
as the event becomes more difficult to observe in direct simulations.
In addition to these probability estimates, we can also analyze the
dominant Gaussian fluctuations around the instanton now and easily sample
high strain events within the Gaussian approximation.
Figure~\ref{fig:nse-evecs} shows the instanton $u_z$ at final time, i.e.\ an
axially symmetric pair of counter-rotating vortex rings, as well as the
dominant eigenfunctions of ${\cal C}_z(T,T)$, corresponding to the fluctuation
modes that are most easily observed at final time in conditioned direct
numerical simulations. 
Note that the Riccati
equation~\eqref{eq:riccati-fw} would be a
PDE for a six-dimensional matrix-valued field $Q_z(x_1,x_2,x_3,y_1,y_2,y_3,t)$
here without
obvious sparsity properties. Solvers for such a problem are quite
expensive, if feasible at all, and also not easy to scale to higher
spatial resolutions, whereas this is possible for the dominant eigenvalue
approach.

\section{Summary and Outlook}

In this paper, we have presented an asymptotically sharp,
sampling-free
probability estimation method for extreme events of stochastic processes
described by additive-noise SDEs and SPDEs.
The method can be regarded as a path-space SORM
approximation. We have introduced and compared two different
conceptual and numerical strategies
to evaluate the pre-exponential factor appearing in these estimates, either
through dominant eigenvalues of the second variation, corresponding to the
standard formulation of precise Laplace asymptotics and SORM, or through
the solution of matrix Riccati differential equations, which is possible for
precise large deviations of continuous-time Markov processes. Highlighting
the scalability of the first approach, we have shown
that leading-order prefactors can be computed in practice
even for very high-dimensional SDEs,
and explicitly tested our methods in two SPDE examples.
In all examples, the approximations showed good agreement with direct Monte
Carlo simulations or importance sampling.
We hope that the methods assembled in this paper are
useful whenever sample path large deviation theory is used to obtain
probability estimates in real-world examples.

There are multiple possible extensions of the methods presented in this
paper. More general classes of SDEs and SPDEs could
possibly be treated numerically within the eigenvalue-based approach,
most notably SDEs with multiplicative Gaussian noise, but also SDEs driven
by Levy noise or singular SPDEs.
Furthermore, one could try to generalize the approach to include
any additive Gaussian noise that is colored in time instead of white.
This would potentially lead to further dimensional reduction for the
instanton and prefactor computation for examples with a slowly
decaying temporal noise correlation. It would also be
interesting to apply the eigenvalue-based prefactor computation strategy to metastable non-gradient SDEs.
Regarding the numerical applicability of the
Riccati method in case of high-dimensional systems with low-rank forcing,
there is an alternative formulation of the prefactor in terms of a
backward-in-time Riccati equation~\citep{grafke-schaefer-vanden-eijnden:2021},
which could be better suited for controlled low-rank approximations.
In general, improvements of the quadratic approximation used throughout
this paper
via loop expansions, resummation techniques or non-perturbative methods
from theoretical physics could be investigated. In this regard, it would
be desirable to obtain simple criteria that indicate whether
the SORM approximation considered in this paper can be expected to
be accurate for given $\eps$ and $z$.
Finally, one could use
the instanton and additional prefactor information for efficient
importance sampling of extreme events for S(P)DEs.

\begin{acknowledgments}
The authors would like to thank Sandra May, Rainer Grauer, and Eric
Vanden-Eijnden for helpful discussions.
T.S.\ acknowledges the support received from the Ruhr University
Research School, funded by Germany's Excellence Initiative [DFG GSC
  98/3], that enabled a research visit at the Courant Institute
of Mathematical Sciences.
T.G. acknowledges the support received from the EPSRC projects
EP/T011866/1 and EP/V013319/1.
\end{acknowledgments}

\appendix

\section{Derivations}
\label{sec:deriv}

\subsection{Laplace Method in finite dimensions}
\label{subsec:deriv-laplace-finite-dim}

In this section, we give a more detailed explanation on how the finite
dimensional Laplace Method is used to estimate extreme event
probabilities in complex systems. It follows arguments similar
to~\cite{dematteis-grafke-vanden-eijnden:2019,
  tong-vanden-eijnden-stadler:2021}.

In
\begin{align}
P_F^\eps(z) =& \left(2 \pi \eps \right)^{-N/2} \times \nonumber\\
 & \times \int_{\RR^N}
\mathds{1}_{\{F \geq z\}}(\eta) \exp \left\{-\frac{1}{2 \eps}
\norm{\eta}_N^2 \right\}  \dd^N \eta\,,
\end{align}
we expand
\begin{align}
\eta = \eta_z + \eps \eta_1 + \sqrt{\eps} \eta_2
\label{eq:eta-scaling}
\end{align}
with $\eta_1, \eta_2 \in \RR^N$ satisfying $\eta_1 \parallel \eta_z$ and $\eta_2 \in \eta_z^\perp$,
such that
\begin{align}
\frac{1}{2 \eps}\norm{\eta}_N^2 = \frac{\eps}{2} \norm{\eta_1
}^2_N + \left \langle \eta_1, \eta_z \right \rangle_N
+ \frac{1}{\eps} I_F(z) + \frac{1}{2} \norm{\eta_2}^2_N
\end{align}
and
\begin{align}
\frac{F(\eta) - z}{\eps} =& \frac{1}{\lambda_z} \left \langle
\eta_1, \eta_z \right \rangle_N \nonumber\\
&+ \frac{1}{2} \left
\langle \eta_2, \nabla^2 F(\eta_z) \eta_2 \right \rangle_N
+ {\cal O}\left(\eps^{1/2} \right)\,.
\end{align}
To motivate the decomposition~\eqref{eq:eta-scaling}, note that
the natural scaling for random fluctuations around the
fixed state $\eta_z$ is clearly $\propto\sqrt{\eps}$, and we use this ansatz
for all directions except for the one parallel to the instanton.
In this direction, due to the restriction $F \geq z$ of the event set,
we can expect a different behavior, and the subsequent computations
in this section confirm that a decay with $\eps$ faster than
$\sqrt{\eps}$ is indeed observed.
We obtain, with $\eta_1 = s e_z$ for $s \in \RR$ and $e_z:= {\eta_z}/{\|\eta_z\|_N}$,
\begin{align}
&P_F^\eps(z) \overset{\eps \downarrow 0}{\sim} (2 \pi)^{-N/2}
\eps^{1/2} \exp \left\{-\eps^{-1} I_F(z) \right\} \times  \nonumber\\
& \quad \times \int_{\eta^\perp_z} \dd^{N-1} \eta_2 \; \exp
\left\{-\frac{1}{2} \norm{\eta_2}^2_{N} \right\} \times \nonumber\\
& \quad\times \int_{-
\frac{\lambda_z}{2 \norm{\eta_z}_N} \left \langle \eta_2,
\nabla^2 F(\eta_z) \eta_2 \right \rangle_N}^\infty \dd s \;
\exp \left\{- s \norm{\eta_z}_N \right\} \nonumber\\
&= (2 \pi)^{-N/2} \eps^{1/2} \norm{\eta_z}_N^{-1} \exp \left\{
-\eps^{-1} I_F(z) \right\} \times \nonumber\\
& \quad\times\int_{\eta^\perp_z} \dd^{N-1} \eta_2
\; \exp \left\{-\frac{1}{2} \left \langle \eta_2, \left( 1_{N
\times N} - \lambda_z \nabla^2 F(\eta_z) \right) \eta_2
\right \rangle_{N} \right\} \nonumber\\
&= (2 \pi)^{-1/2} \eps^{1/2} \exp \left\{-\eps^{-1}
I_F(z) \right\} \times \nonumber \\
&\quad \times \left[2 I_F(z)\det \left(1_{N
\times N} - \lambda_z \, \ppr_{\eta_z^\perp} \nabla^2 F(\eta_z)
\ppr_{\eta_z^\perp} \right) \right]^{-1/2}\,.
\end{align}
With this computation, we have motivated~\eqref{eq:extreme-event-finite-dimensional-result}
and~\eqref{eq:extreme-event-finite-dimensional-prefac}. A rigorous
proof would consist of a more careful error analysis for the
Laplace method, as detailed e.g.\ by~\cite{bleistein:1975}.

\subsection{Laplace Method in infinite dimensions}
\label{subsec:deriv-laplace-infinite-dim}

It is a common strategy in large deviation theory to first study
expectations of the type $\EE \left[\exp \left\{\frac{1}{\eps}
  F(\phi^\eps) \right\} \right]$ for a family of random variables
$\phi^\eps$ satisfying a large deviation principle, and a real-valued
function $F$. Only later will these results be transformed onto
probabilities or other probabilistic quantities.  We directly use
the results of~\cite{ben-arous:1988} to conclude that the asymptotic
behavior of the moment-generating function (MGF) $A_F^\eps \colon \RR
\to [0, \infty]$, $A_F^\eps(\lambda) = \EE \left[ \exp
  \left\{\tfrac{\lambda}{\eps} f(X_T^\eps) \right\}\right]$ of the
observable $f(X^\eps_T)$ for the additive-noise SDE~\eqref{eq:sde} as
$\eps \downarrow 0$ is given by
\begin{align}
 A_F^\eps(\lambda)\overset{\eps \downarrow 0}{\sim} R_\lambda
    \exp \left\{\eps^{-1} I_F^*(\lambda) \right\}
    \label{eq:mgf-general}
\end{align}
  with prefactor
  \begin{align}
  R_\lambda = \left[\det\left(\Id - \left. \lambda
  \nfdv{2}{F}{\eta} \right|_{\eta_\lambda} \right) \right]^{-1/2}\,.
  \label{eq:ratio-funcdet}
  \end{align}
Here, $I_F^*$ denotes the Legendre transform of the rate function
$I_F$, $\det$ is a Fredholm determinant, the second variation
operator $\left. \nfdv{2}{F_\lambda}{\eta} \right|_{\eta_\lambda}$ is
trace class, and $\eta_\lambda$ is short for $\eta_{z_\lambda}$ at
the Legendre dual $z_\lambda$ of $\lambda$ via $I_F'(z_\lambda) =
\lambda$. Note that for multiplicative noise, the result would
be different, which can already be seen in the simple example of a
one-dimensional geometric Brownian motion and $f(x) = \tfrac{1}{2}
\log^2 x$. Furthermore, \cite{ben-arous:1988} also assumes
that the vector field $b$, the observable $f$, and their respective
derivatives are bounded.
A remark by~\cite{deuschel-etal:2014} shows how one could relax
this assumption via localization.

Evaluating
the inverse Laplace transform from the MGF~\eqref{eq:mgf-general}
to the probability
density function
\begin{align}
\rho_F^\eps(z) &= \frac{1}{2\pi i \eps} \int_C A_F^\eps(\lambda)
\exp\left\{-\frac{\lambda z}{\eps}
\right\}\;\dd \lambda \nonumber\\
&\overset{\eps \downarrow 0}{\sim} \left(2 \pi \eps \right)^{-1/2}
R_{\lambda_z} \sqrt{I_F''(z)} \exp \left\{- \eps^{-1} I_F(z) \right\}
\end{align}
via a saddlepoint approximation, as well as a further integration
to get the tail probability via a Laplace approximation yields the
desired estimate with leading-order prefactor
\begin{align}
C_F(z) = R_{\lambda_z}\sqrt{I_F''(z)}\lambda_z^{-1}\,.
\label{eq:pref-rlbda}
\end{align}
From the first-order necessary condition
\begin{align}
\eta_z = \lambda_z \left. \fdv{F}{\eta} \right|_{\eta_z}
\end{align}
and $\lambda_z = I_F'(z)$, we get via differentiation
\begin{align}
\frac{\lambda_z}{I_F''(z)} \dv{\eta_z}{z} = \left[\Id - \lambda_z\left.
  \nfdv{2}{F}{\eta} \right|_{\eta_z} \right]^{-1} \eta_z\,,
\end{align}
so
\begin{align}
C_F(z) &= \left[\left \langle \eta_z, \frac{\lambda_z}{I_F''(z)}
\dv{\eta_z}{z} \right \rangle_{L^2} \det\left(\Id -
\left. \lambda_z
  \nfdv{2}{F}{\eta} \right|_{\eta_z} \right) \right]^{-1/2} \nonumber\\
  &= \left[2 I_F(z) \left \langle \frac{\eta_z}{\norm{\eta_z}},
  \left[\Id - \lambda_z\left.
  \nfdv{2}{F}{\eta} \right|_{\eta_z} \right]^{-1} \frac{\eta_z
  }{\norm{\eta_z}}\right \rangle_{L^2} \right. \times \nonumber\\
  &\quad \times \left. \det\left(\Id
  - \left. \lambda_z
  \nfdv{2}{F}{\eta} \right|_{\eta_z} \right) \right]^{-1/2} \nonumber\\
  &=  \left[2 I_F(z) \det \left(\Id - \lambda_z
\ppr_{\eta_z^\perp} \left. \nfdv{2}{F}{\eta} \right|_{
\eta_z} \ppr_{
\eta_z^\perp} \right) \right]^{-1/2}
\end{align}
as claimed. The last equality is easy to see for finite-dimensional
matrices: For $A \in \RR^{N \times N}$ invertible and a unit vector $e \in
\RR^N$, the adjugate is $\text{adj}(A) = \det A \cdot A^{-1}$, and
applying $e$ from the left and right yields $\det A \left \langle e,
A^{-1} e \right \rangle_N = \left \langle e, \text{adj}(A) e \right
\rangle_N$. The right-hand side is the $(e,e)$ cofactor of $A$, which
is equal to the determinant of the $(N-1) \times (N-1)$ matrix $
\ppr_{e^\perp} A \ppr_{e^\perp}$ with $\ppr_{e^\perp}
\colon \RR^N \to e^\perp$ denoting the orthogonal projection. For
the present infinite-dimensional case, an analogue of this relation
can be verified using the series definition of the Fredholm determinant
and adjugate as originally introduced by Fredholm
himself~\citep{fredholm:1903,mckean:2011}.

\subsection{From Fredholm determinants to zeta-regularized functional determinants}
\label{subsec:deriv-fred-zeta}
In this section, we motivate~\eqref{eq:det-ratio} using purely formal
manipulations, and only consider linear observables $f$ for
simplicity. For rigorous results on the relation between Fredholm
determinants and zeta-regularized determinants for related classes of
operators, see e.g.~\cite{forman:1987,hartmann-lesch:2022}.  We start
with the expression~\eqref{eq:pref-rlbda} for the prefactor~$C_F(z)$
in terms of the full second variation determinant without projection
operators.  According to the adjoint formulation of
section~\ref{subsec:pref-fred-adj}, we write the second variation
$\delta^2 (\lambda F) / \delta \eta^2$ as the composition of three
linear operators
\begin{align}
\lambda_z \left.\nfdv{2}{F}{\eta} \right|_{\eta_z} = \left[ L^\top_{
z,(T,0)} \right]^{-1} \circ \left \langle \nabla^2 b(\phi_z),
\theta_z \right \rangle_n \circ \left[ L_{z,(0,0)} \right]^{-1}\,.
\end{align}
Here, the operator in the middle simply denotes pointwise multiplication
with $\left \langle \nabla^2 b(\phi_z(t)), \theta_z(t)
\right \rangle_n$ for each $t \in [0,T]$. The rightmost operator, for a given
argument $\delta \eta$, integrates
\begin{align}
\begin{cases}
\dot{\gamma} = \nabla b(\phi_z) \gamma + \sigma \delta \eta\,, \\
\gamma(0) = 0
\end{cases}
\end{align}
and sets $\left[ L_{z,(0,0)} \right]^{-1} \delta \eta = \gamma$.
Symbolically, we have
\begin{align}
\left[ L_{z,(0,0)} \right]^{-1} = \left[\dv{}{t} - \nabla
b(\phi_z) \right]^{-1}_{(0,0)} \sigma,
\end{align}
where the subscript denotes inversion under the boundary condition
$\gamma(0)=0$. Similarly, we put $\left[ L_{z,(T,0)}^\top \right]^{-1}
\gamma = \zeta$
with
\begin{align}
 \left[ L^\top_{z,(T,0)} \right]^{-1} = \sigma^\top \left[-\dv{}{t} -
 \nabla b^\top(\phi_z) \right]^{-1}_{(T,0)},
\end{align}
under the boundary condition $\zeta(T) = 0$. Symbolically, we then get
\begin{align}
&\quad \left[\det\left(\Id - \lambda_z \left.
  \nfdv{2}{F}{\eta} \right|_{\eta_z} \right) \right]^{-1/2} \nonumber\\
  &=
  \left[\det\left(\Id - \left[ L^\top_{
z,(T,0)} \right]^{-1} \circ \left \langle \nabla^2 b(\phi_z),
\theta_z \right \rangle_n \circ \left[ L_{z,(0,0)} \right]^{-1}
\right) \right]^{-1/2} \nonumber \\
&= \left[ \frac{\Det \left(L^\top_{
z,(T,0)}  L_{z,(0,0)} - \left \langle \nabla^2 b(\phi_z),
\theta_z \right \rangle_n  \right)}{\Det \left(L^\top_{
z,(T,0)}  L_{z,(0,0)} \right)} \right]^{-1/2} \nonumber\\
&=
\left[ \frac{\Det \left(L^\top_{
z,(T,0)}  L_{z,(0,0)} - \left \langle \nabla^2 b(\phi_z),
\theta_z \right \rangle_n  \right)}{\Det \left(L^\top_{
0,(T,0)}  L_{0,(0,0)} \right)} \right]^{-1/2} \times \nonumber \\
& \quad \times \left[ \frac{\Det \left(L^\top_{
0,(T,0)}  L_{0,(0,0)} \right)}{\Det \left(L^\top_{
z,(T,0)}  L_{z,(0,0)} \right)} \right]^{-1/2}\,.
\label{eq:det-reformulation}
\end{align}
Here, the critical step is in the second line where
the operators are moved out of the Fredholm determinant to get a fraction of
two zeta-regularized determinants, which is true for finite-dimensional matrices
but non-trivial for general operators. We see that the
boundary conditions of all appearing operators are
\begin{align}
{\cal A}_0\colon \begin{cases}
\gamma(0) = 0\,,\\
\zeta(T)=0\,,
\end{cases}
\end{align}
which is the correct special case of the general boundary conditions
\begin{align}
{\cal A}_{\lambda_z}\colon \begin{cases}
\gamma(0) = 0\,,\\
\zeta(T)= \lambda_z \nabla^2 f(\phi_z(T)) \gamma(T)
\end{cases}
\end{align}
from~\cite{schorlepp-grafke-grauer:2023} for a linear observable $f$.
Moreover, we have
\begin{align}
&L^\top_{
z,(T,0)}  L_{z,(0,0)} - \left \langle \nabla^2 b(\phi_z),
\theta_z \right \rangle_n \nonumber \\
&= \left[-\dv{}{t}
     - \nabla b^\top(\phi_z) \right] \ a^{-1} \left[\dv{}{t} -
     \nabla b(\phi_z) \right] - \langle \nabla^2 b(\phi_z),\theta_z
     \rangle_n \nonumber\\
     &=: \Omega[\phi_z] \,,
\end{align}
which is the Jacobi operator, defined via $\delta^2 S[\phi_z][\gamma] =
\frac{1}{2} \int_0^T \left \langle \gamma, \Omega[\phi_z] \gamma
\right \rangle_n \dd t$, for the Freidlin-Wentzell action functional
$S[\phi] = \tfrac{1}{2} \int_0^T \langle \dot{\phi}
- b(\phi), a^{-1}(\dot{\phi} - b(\phi) ) \rangle_n \, \dd t$.
We then use Forman's theorem~\citep{forman:1987} to evaluate the
second ratio of determinants in~\eqref{eq:det-reformulation}
\begin{align}
 &\left[ \frac{\Det \left(L^\top_{
0,(T,0)}  L_{0,(0,0)} \right)}{\Det \left(L^\top_{
z,(T,0)}  L_{z,(0,0)} \right)} \right]^{-1/2} \nonumber\\
&\quad = 
\exp\left\{-\tfrac12 \int_0^T \left(\nabla\cdot
  b(\phi_z) - \nabla\cdot b(\phi_0)\right)\,\dd t\right\}\,,
\end{align}
thereby finishing the motivation of the result~\eqref{eq:det-ratio}.

\subsection{Full covariance function via eigenvalues and eigenfunctions}
\label{subsec:deriv-eig-cov}
In this section, we formally derive~\eqref{eq:cov-eig}. First, we
introduce the evaluation maps $\Phi_t$ for $t \in [0,T]$ as
$(\eta(s))_{s \in [0,T]} \xrightarrow{\Phi_t} \phi(t)$ with
\begin{align}
\begin{cases}
\dot{\phi} = b(\phi) + \sigma \eta\,,\\
\phi(0) = x\,.
\end{cases}
\end{align}
Then
\begin{align}
&{\cal C}_z(t,t') \nonumber\\
&= \lim_{\eps \downarrow 0} \EE \left[\frac{(X_t^\eps -
\phi_z(t))\otimes(X_{t'}^\eps -
\phi_z(t'))}{\eps} \bigg\mid f(X_T^\eps) = z\right] \nonumber\\
&= \lim_{\eps \downarrow 0} \left(\eps \EE \left[\delta(f(\Phi_T[\sqrt{\eps}\eta]) - z) \right] \right)^{-1} \times \nonumber\\
& \quad \times \left( \EE \left[(\Phi_t[\sqrt{\eps}\eta] -
\phi_z(t))\otimes(\Phi_{t'}[\sqrt{\eps}\eta] -
\phi_z(t')) \right. \right. \times \nonumber\\
& \quad \times \left. \left. \delta(f(\Phi_T[\sqrt{\eps}\eta]) - z) \right] \right),
\label{eq:cov-PI}
\end{align}
where $\delta$ denotes the Dirac delta function. The denominator
of~\eqref{eq:cov-PI} is
just the PDF $\rho_F^\eps(z)$; we already know its asymptotic behavior
from~\eqref{eq:pdf-estimate}. In short, its asymptotics are obtained as
\begin{align}
&\EE \left[\delta(f(\Phi_T[\sqrt{\eps}\eta]) - z) \right] \nonumber\\
&=\frac{1}{2 \pi i \eps} \int_{-i \infty}^{i \infty} \dd \lambda
\int D \eta \, \exp \left\{-\frac{1}{\eps} \left(
\frac{1}{2}\norm{\eta}^2_{L^2} - \lambda (F[\eta] - z)
\right) \right\} \nonumber \\
&\overset{\eps \downarrow 0}{\sim} \frac{1}{2 \pi i \eps^{1/2}}
\exp \left\{-I_F(z) / \eps \right\} \int_{-i \infty}^{i \infty}
\dd \lambda \int D \eta \times \nonumber\\
&\times\exp \left\{-\frac{1}{2} \left \langle
\eta, \left[\Id - \lambda_z \left. \nfdv{2}{F}{\eta} \right|_{\eta_z}
\right] \eta \right \rangle_{L^2} \right\}
\exp \bigg\{ \lambda\bigg \langle \underbrace{\left. \fdv{F}{\eta}
\right|_{\eta_z}}_{= \eta_z / \lambda_z}, \eta \bigg \rangle_{L^2}
\bigg\} \nonumber\\
&= \frac{1}{\eps^{1/2}} \exp \left\{-I_F(z) / \eps \right\}
\frac{\abs{\lambda_z}}{\norm{\eta_z}_{L^2}} \times\nonumber\\
&\times \underbrace{\int D \eta \,
\exp \left\{-\frac{1}{2} \left \langle \eta, \left[\Id - \lambda_z \left.
\nfdv{2}{F}{\eta} \right|_{\eta_z}
\right] \eta \right \rangle_{L^2} \right\} \delta \left( \left
\langle e_z, \eta \right \rangle_{L^2} \right)}_{=(2 \pi)^{-1/2}
\det \left(\Id - A_z \right)^{-1/2}}\,.
\end{align}
Here, in the first step, the PDF was written as the inverse
Laplace transform of the moment-generating function, and the
expectation over $\eta$ was expressed as a functional integral.
Then, in the second step, all integration variables were expanded
up to second order around the stationary point $(\eta_z, \lambda_z)$.
Finally, in the last step, the $\lambda$ integral was interpreted as
a delta function again, restricting the functional integration to
the subspace orthogonal to $e_z = \eta_z / \norm{\eta_z}_{L^2}$.
Hence, the Gaussian integral yields the determinant in the
subspace $\eta_z^\perp$, and the factor of $(2 \pi)^{-1/2}$ appears
due to the normalization of the functional integral.
For the numerator of~\eqref{eq:cov-PI}, we proceed similarly:
\begin{align}
&\eps^{-1} \EE \left[(\Phi_t[\sqrt{\eps}\eta] -
\phi_z(t))\otimes(\Phi_{t'}[\sqrt{\eps}\eta] -
\phi_z(t')) \times \right. \nonumber\\
&\qquad \quad  \times\left. \delta(f(\Phi_T[\sqrt{\eps}\eta]) - z) \right] \nonumber\\
&\overset{\eps \downarrow 0}{\sim} \frac{1}{\eps^{1/2}}
\exp \left\{-I_F(z) / \eps \right\}
\frac{\abs{\lambda_z}}{\norm{\eta_z}_{L^2}} \times \nonumber\\
&\quad \times  \int D \eta \,
\exp \left\{-\frac{1}{2} \left \langle \eta, \left[\Id - \lambda_z \left.
\nfdv{2}{F}{\eta} \right|_{\eta_z}
\right] \eta \right \rangle_{L^2} \right\} \times \nonumber\\
& \quad \times\delta \left( \left
\langle e_z, \eta \right \rangle_{L^2} \right) \left. \delta
\Phi_t \right|_{\eta_z}[\eta] \otimes  \left. \delta \Phi_{t'}
\right|_{\eta_z}[\eta]\,,
\end{align}
where $\left. \delta \Phi_t \right|_{\eta_z}$ denotes the
first variation of $\Phi_t$. One can show, by first using an
adjoint variable and then proceeding similar to
the boundary condition computation in
section~\ref{sec:deriv}\ref{subsec:deriv-bc}, that
\begin{align}
\left. \delta
\Phi_t \right|_{\eta_z}[\eta] = \gamma(t)
\end{align}
is the state space fluctuation from~\eqref{eq:second-order-adj-eq}
around $\phi_z$ at time~$t$ associated with~$\eta$.
Since this is a linear function of~$\eta$, expanding
\begin{align}
\eta = \sum_{i = 1}^\infty \alpha_i \delta \eta_z^{(i)}
\end{align}
in terms of the orthonormal eigenfunctions of $A_z$ and performing
the Gaussian integration in the $\alpha$ variables then leads
to~\eqref{eq:cov-eig}.

\subsection{Final time conditioned fluctuations boundary condition}
\label{subsec:deriv-bc}
Here, we show that for the state variable fluctuations $\gamma$ associated
with any $\delta \eta \in \eta_z^\perp \subset L^2([0,T],\RR^n)$, the 
final time boundary condition
\begin{align}
\left \langle \lambda_z \nabla f(\phi_z(T)), \gamma(T) \right \rangle_n = 0
\end{align}
holds, and hence the result~\eqref{eq:cov-eig} for the fluctuation covariance
in terms of the $\gamma^{(i)}_z$'s is consistent with~\eqref{eq:final-bc}.
Note that the linearized state equation for $\gamma$
in~\eqref{eq:second-order-adj-eq} can be formally integrated to get
\begin{align}
\gamma(T) = \int_0^T {\cal T} \left[ \exp \left\{ \int_t^T
\nabla b(\phi_z(\tau)) \, \dd \tau \right\} \right] \sigma
\delta \eta(t) \, \dd t\,,
\end{align}
where ${\cal T}$ is the time-ordering operator. Similarly, from the 
first order adjoint equation in~\eqref{eq:first-order-adj-eq}, we get
\begin{align}
\theta_z(t) = {\cal T} \left[ \exp \left\{  \int_t^T \nabla b(\phi_z(\tau))^\top
\dd \tau \right\} \right] \lambda_z \nabla f(\phi_z(T))\,,
\end{align}
and hence
\begin{align}
\left \langle \lambda_z \nabla f(\phi_z(T)), \gamma(T) \right
\rangle_n = \left \langle \eta_z, \delta \eta \right \rangle_{L^2} = 0\
\end{align}
by transposing.

\newpage
\bibliography{bib}
\end{document}